\documentclass[prd,twocolumn,floatfix,amsmath,nofootinbib,amssymb,floatfix]{revtex4}
\usepackage{graphicx,color,dcolumn,booktabs,bm}
\usepackage{longtable,lscape}
\usepackage{pdfpages}
\usepackage{txfonts}
\usepackage{overpic}
\usepackage{amssymb}
\usepackage{makecell}
\usepackage{indentfirst}
\usepackage{feynmf}   
\usepackage{slashed}  
\usepackage{cases}
\usepackage{color}
\usepackage{multirow}
\usepackage{threeparttable}
\usepackage{epstopdf}
\usepackage{enumerate}
\usepackage{subfigure}
\usepackage{diagbox}
\usepackage{graphicx,color,dcolumn,booktabs,bm}
\usepackage{mathrsfs}
\usepackage{cancel}
\usepackage{float}
\usepackage[colorlinks,
            citecolor=blue,
            anchorcolor=red,
            menucolor=red,
            linkcolor=red,
            filecolor=red,
            runcolor=red,
            urlcolor=blue,
            frenchlinks=red]{hyperref}

\begin{document}
\title{The ground states of hidden-charm tetraquarks and their radial excitations}
\author{Guo-Liang Yu$^{1}$}
\email{yuguoliang2011@163.com}
\author{Zhen-Yu Li$^{2}$}
\email{zhenyvli@163.com}
\author{Zhi-Gang Wang$^{1}$}
\email{zgwang@aliyun.com}
\author{Bin Wu$^{1}$}
\author{Ze Zhou$^{1}$}
\author{Jie Lu$^{1}$}

\affiliation{$^1$ Department of Mathematics and Physics, North China
Electric Power University, Baoding 071003, People's Republic of
China\\$^2$ School of Physics and Electronic Science, Guizhou Education University, Guiyang 550018, People's Republic of
China}
\date{\today }

\begin{abstract}
Inspired by the great progress in the observations of charmonium-like states in recent years, we perform a systematic analysis about the ground states and the first radially excited states of $qc\bar{q}\bar{c}$ ($q$=$u/d$ and $s$) tetraquark systems. Their mass spectra, root mean square (r.m.s.) radii and radial density distributions are predicted within the framework of relativized quark model. By comparing with experimental data, some potential candidates for hidden-charm tetraquark states are suggested. For $qc\bar{q}\bar{c}$ ($q$=$u/d$) system, if $Z_{c}(3900)$ is supposed to be a compact tetraquark state with $J^{PC}=1^{+-}$, $Z(4430)$ can be interpreted as the first radially excited states of $Z_{c}(3900)$.
Another broad structure $Z_{c}(4200)$ can also be explained as a partner of $Z_{c}(3900)$, and it arise from a higher state with $J^{PC}=1^{+-}$. In addition, theoretical predictions indicate that the possible assignments for $X(3930)$, $X(4050)$ and $X(4250)$ are low lying $0^{++}$ tetraquark states. As for the $sc\bar{s}\bar{c}$ system, $X(4140)$ and $X(4274)$ structures can be interpreted as this type of tetraquark states with $J^{PC}=1^{++}$, and $X(4350)$ can be described as a $sc\bar{s}\bar{c}$ tetraquark with $J^{PC}=0^{++}$. With regard to $qc\bar{s}\bar{c}$ ($q$=$u/d$) system, we find two potential candidates for this type of tetraquark, which are $Z_{cs}(4000)$ and $Z_{cs}(4220)$ structures. The measured masses of these two structures are in agreement with theoretical predictions for the $1^{+}$ state.
\end{abstract}


\maketitle

\section{Introduction}\label{sec1}

When the quark model \cite{qmodel1,qmodel2} was first proposed, it predicted not only the existence of the mesons($q\bar{q}$) and baryons($qqq$) but also the existence of multi-quark systems such as tetraquarks and pentaquarks. Since then, scientists devoted much efforts to searching for the multi-quark states in experiments. However, the experimental results were not satisfactory at first and fewer candidates for multi-quark states were discovered. The breakthrough came in 2003, the famous exotic state $X(3872)$, which is a good candidate for tetraquark state, was first observed by Belle Collaboration \cite{Belle3872}. After that, more and more charmonium-like states emerged like bamboo shoots after a spring rain. The appearance of these exotic states has motivated theorists to devote more energies in studying the nature of these states. In general, these new observed states can not be categorized as the conventional mesons or baryons, and they are commonly explained as compact tetraquark states, hadronic molecular states or admixtures of them. Among these interpretations, the hidden-charm tetraquarks which are composed of two charmed quarks and two light quarks ($u$, $d$ or $s$ quark) are especially interesting. These tetraquarks have rich spectra structures and they are natural laboratories for us to study the interactions between quarks.

Although the nature of $X(3872)$ is still controversial, this can not stop people's enthusiasm for searching for multi-quark states in experiments.
In the successive years after the observation of $X(3872)$, more charmonium-like states such as $Z(3930)$ \cite{Belle3930}, $Z(4430)$ \cite{Belle4430}, $Z(4050)$ \cite{Belle4050}, $Z(4250)$ \cite{Belle4050}, $Y(4140)$ \cite{CDF4140}, $X(4274)$ \cite{CDF4274}, $X(3915)$ \cite{Belle3915}, $Z_{c}(3900)$ \cite{BESIII3900} and $Z_{c}(4020)$ \cite{BESIII4020} were discovered by Belle, CDF, BESIII and LHCb Collaborations. Especially, the charged charmonium-like states $Z_{c}(3900)$, $Z_{c}(4020)$, $Z(4430)$, $Z(4050)$ and $Z(4250)$ may be genuine exotic hadrons with four-quark contents. Furthermore, in recent years, a few charged charmonium-like states with strangeness such as $Z_{cs}(3985)$ \cite{BESIII3985}, $Z_{cs}(4000)$ \cite{LHCb4000} and $Z_{cs}(4220)$ \cite{LHCb4000} were also observed. They are also good candidates for tetraquark states.

From theoretical sides, scientists employed many methods/models to carry out their studies on the structure, the production, and the decay behavior of the exotic states such as various quark model \cite{Maiani:2004vq,Yang:2009zzp,Ortega:2009hj,Yang:2010sf,Deng:2016rus,Richard:2017una,Deng:2017xlb,Luo:2017eub,
Deng:2014gqa,Xiao:2019spy,Hao:2019fjg,Giron:2020qpb,Lebed:2016yvr,Liu:2021xje,Duan:2020tsx,JinX 2021,Yang:2021zhe,Yang:2023mov,Kanwal:2022ani,Wang:2022clw,Lebed:2022vks,Meng:2023jqk,Bayar:2022dqa,Huang:2023jec,Li:2023wxm,Wang:2024vjc,Zhang:2022qtp,Liu:2022hbk,Wang:2021kfv,Wang:2022yes}, QCD sum rule \cite{Chen:2013pya,Chen:2015ata,Agaev:2017tzv,Wang:2013vex,Wang:2017jtz,Wang:2017dtg,Wang:2018ntv,Wang:2020dgr,Wang:2019tlw,Wang:2023jaw,Wang:2024ciy,Agaev:2020zad,Azizi:2020itk,
Chen:2021erj,Yang:2022zxe,Mutuk:2022ckn,Agaev:2023ara}, the effective field theory \cite{Liu:2009qhy,WangP 2013,HeJ 2013,Aceti:2014kja,Ding:2020dio,Wang:2020dya,ChenR 2021,Feijoo:2021ppq,Wang:2022ztm,Xie:2022lyw,
Wu:2023fyh,Chen:2023eix,Qiu:2023uno,Yang:2021sue}, Lattice QCD\cite{Liu:2019gmh,Bicudo:2019mny}, Bethe-Salpeter equation \cite{Wallbott:2019dng,Wallbott:2020upx,Li:2021ygk} and others \cite{Ferretti:2020civ,Ortega:2020lhr,Giacosa:2019zxw,Giron:2020fvd,Duan:2022upr,Guo:2022zbc,Ji:2022blw,Li:2022fgd,Badalian:2022hfu,Huang:2022zsy,
Zhang:2020rqr,Badalian:2023qyi}. One can also consult the reviews \cite{Chen:2016qju,Lebed:2016hpi,Guo:2017jvc,Liu:2019zoy,Brambilla:2019esw,Dong:2021bvy} for more experimental data and theoretical studies about the exotic hadrons. However, interpretations of different methods/models about the structure of charmonium-like states do not agree well with each other, which needs further comparisons and confirmations in theory. This constitutes the first motivation of our present work. Besides, within the framework of quark model, the previous studies mainly concentrated on the non-relativistic quark potential models or simple quark models. As for the hidden-charm tetraquark systems which include two light quarks, the relativistic corrections for the mass spectra should be significant. Therefore, it is necessary and interesting to perform a systematic analysis of the mass spectra of hidden-charm tetraquark systems using the relativized quark model.

The relativized quark model which was first proposed by Godfrey, Capstick and Isgur \cite{Godfrey:1985xj,Capstick:1986ter} has been widely adopted to study the properties of the mesons, baryons, and tetraquarks states \cite{Lu:2021kut,Lu:2017meb}. In our previous works, we used this model to study the mass spectra, r.m.s. radii and the radial density distributions of the singly and doubly charmed baryons \cite{Yu:2022ymb,Li:2022xtj,Yu:2022lel,Li:2022oth}, and the fully charmed tetraqark states \cite{Yu:2022lak}. In the present work, we extend our previous method to analyze the hidden-charm tetraquark states. We hope this study can help to shed more light on the nature of these charmonium-like states. The paper is organized as follows. After the introduction, we briefly describe the phenomenological method adopted in this work in Sec. \ref{sec2}. In Sec. \ref{sec3} we present our numerical results. Sec. \ref{sec4} is devoted to discuss the mass spectra in detail and try to find potential candidates for hidden-charm tetraquark states, Sec. \ref{sec5} is reserved for our conclusions.

\section{Phenomenological method adopted in this work}\label{sec2}

Within the framework of relativized quark model, its Hamiltonian for a four-body system includes a relativistic kinetic energy term, a confining potentials and the one-gluon exchange potential \cite{Lu:2021kut,Lu:2017meb},
\begin{eqnarray}
H=\sum_{i=1}^{4}(p_{i}^{2}+m_{i}^{2})^{1/2}+\sum_{i<j}V_{ij}^{\mathrm{conf}}+\sum_{i<j}V_{ij}^{\mathrm{oge}}
\end{eqnarray}
$V_{ij}^{\mathrm{conf}}$ is the confining potential which can be written as,
\begin{eqnarray}
\notag
V_{ij}^{\mathrm{conf}}&&=-\frac{3}{4}\textbf{\emph{F}}_{i}\cdot\textbf{\emph{F}}_{j}\Big[b r_{ij}\big[\frac{e^{-\sigma_{ij}^{2}r_{ij}^{2}}}{\sqrt{\pi}\sigma_{ij} r_{ij}}\\
&&+\big(1+\frac{1}{2\sigma_{ij}^{2}r_{ij}^{2}}\big)\frac{2}{\sqrt{\pi}} \int^{\sigma_{ij} r_{ij}}_{0}e^{-x^{2}}dx\big]+c\Big]
\end{eqnarray}
with
\begin{eqnarray}
\sigma_{ij}=\sqrt{s^{2}\Big[\frac{2m_{i}m_{j}}{m_{i}+m_{j}}\Big]^{2}+\sigma_{0}^{2}\Big[\frac{1}{2}\big(\frac{4m_{i}m_{j}}{(m_{i}+m_{j})^{2}}\big)^{4}+\frac{1}{2}\Big]}
\end{eqnarray}
$\textbf{\emph{F}}_{i}\cdot\textbf{\emph{F}}_{j}$ stands for the color matrix and $\textbf{\emph{F}}$ reads
\begin{equation}
F_{n}=\left\{
      \begin{array}{l}
       \frac{\lambda_{n}}{2} \quad \mathrm{for} \, \mathrm{quarks}, \\
        -\frac{\lambda_{n}^{*}}{2} \quad    \mathrm{for} \, \mathrm{antiquarks} \\
      \end{array}
      \right.
\end{equation}
with $n=1,2\cdots8$.

Because we only concentrate on the S-wave tetraquark states in this work, the one-gluon
exchange potential $V_{ij}^{\mathrm{oge}}$ includes a Coulomb term $V_{ij}^{\mathrm{Coul}}$ and hyperfine interaction $V^{\mathrm{hyp}}_{ij}$,
\begin{eqnarray}
V^{\mathrm{oge}}_{ij}=V_{ij}^{\mathrm{Coul}}+V^{\mathrm{hyp}}_{ij}
\end{eqnarray}
The Coulomb term $V_{ij}^{\mathrm{Coul}}$ can be expressed as,
\begin{eqnarray}
V_{ij}^{\mathrm{Coul}}=\Big(1+\frac{p^{2}_{ij}}{E_{i}E_{j}}\Big)^{\frac{1}{2}}\widetilde{G}(r_{ij})\Big(1+\frac{p^{2}_{ij}}{E_{i}E_{j}}\Big)^{\frac{1}{2}}
\end{eqnarray}
where $\widetilde{G}(r_{ij})$ is a smeared one-gluon exchange propagator,
\begin{eqnarray}
\widetilde{G}(r_{ij})=\textbf{\emph{F}}_{i}\cdot\textbf{\emph{F}}_{j}\mathop{\sum}\limits_{k=1}^{3}\frac{2\alpha_{k}}{3\sqrt{\pi}r_{ij}}\int^{\tau_{k}r_{ij}}_{0}e^{-x^{2}}dx
\end{eqnarray}
with $\tau_{k}=\frac{1}{\sqrt{\frac{1}{\sigma_{ij}^{2}}+\frac{1}{\gamma_{k}^{2}}}}$.

As for the hyperfine interaction $V^{\mathrm{hyp}}_{ij}$, it is composed of tensor interaction and the contact interaction,
\begin{eqnarray}
V^{\mathrm{hyp}}_{ij}=V^{\mathrm{tens}}_{ij}+V_{ij}^{\mathrm{cont}}
\end{eqnarray}
with
\begin{eqnarray}\label{tens}
\notag
V^{\mathrm{tens}}_{ij}&&=-\frac{1}{3m_{i}m_{j}}\Big(\frac{3\textbf{S}_{i}\cdot \textbf{r}_{ij}\textbf{S}_{j}\cdot \textbf{r}_{ij}}{r_{ij}^{2}}-\textbf{S}_{i}\cdot\textbf{S}_{j}\Big)\\
&&\times\Big(\frac{\partial^{2}}{\partial r_{ij}^{2}}-\frac{1}{r_{ij}}\frac{\partial}{\partial r_{ij}}\Big)\widetilde{G}_{ij}^{\mathrm{t}},
\end{eqnarray}
\begin{eqnarray}\label{cont}
V^{\mathrm{cont}}_{ij}=\frac{2\textbf{S}_{i}\cdot\textbf{S}_{j}}{3m_{i}m_{j}}\bigtriangledown^{2}\widetilde{G}_{ij}^{\mathrm{c}}
\end{eqnarray}
In Eqs. \ref{tens} and \ref{cont}, $\widetilde{G}^{\mathrm{t}}_{ij}$ and $\widetilde{G}^{\mathrm{c}}_{ij}$ are also achieved from the $\widetilde{G}(r_{ij})$ by introducing momentum-dependent factors,
\begin{eqnarray}
\widetilde{G}^{\mathrm{t}}_{ij}=\Big(\frac{m_{i}m_{j}}{E_{i}E_{j}}\Big)^{\frac{1}{2}+\epsilon_{\mathrm{t}}}\widetilde{G}(r_{ij})\Big(\frac{m_{i}m_{j}}{E_{i}E_{j}}\Big)^{\frac{1}{2}+\epsilon_{\mathrm{t}}}
\end{eqnarray}
\begin{eqnarray}
\widetilde{G}^{\mathrm{c}}_{ij}=\Big(\frac{m_{i}m_{j}}{E_{i}E_{j}}\Big)^{\frac{1}{2}+\epsilon_{\mathrm{c}}}\widetilde{G}(r_{ij})\Big(\frac{m_{i}m_{j}}{E_{i}E_{j}}\Big)^{\frac{1}{2}+\epsilon_{\mathrm{c}}}
\end{eqnarray}
with $E_{i}=\sqrt{m_{i}^{2}+p_{ij}^{2}}$, where $p_{ij}$ is the magnitude of the momentum of either of the quarks in the $ij$ center-of-mass frame.

\begin{figure}[h]
\centering
\includegraphics[height=3.5cm,width=8cm]{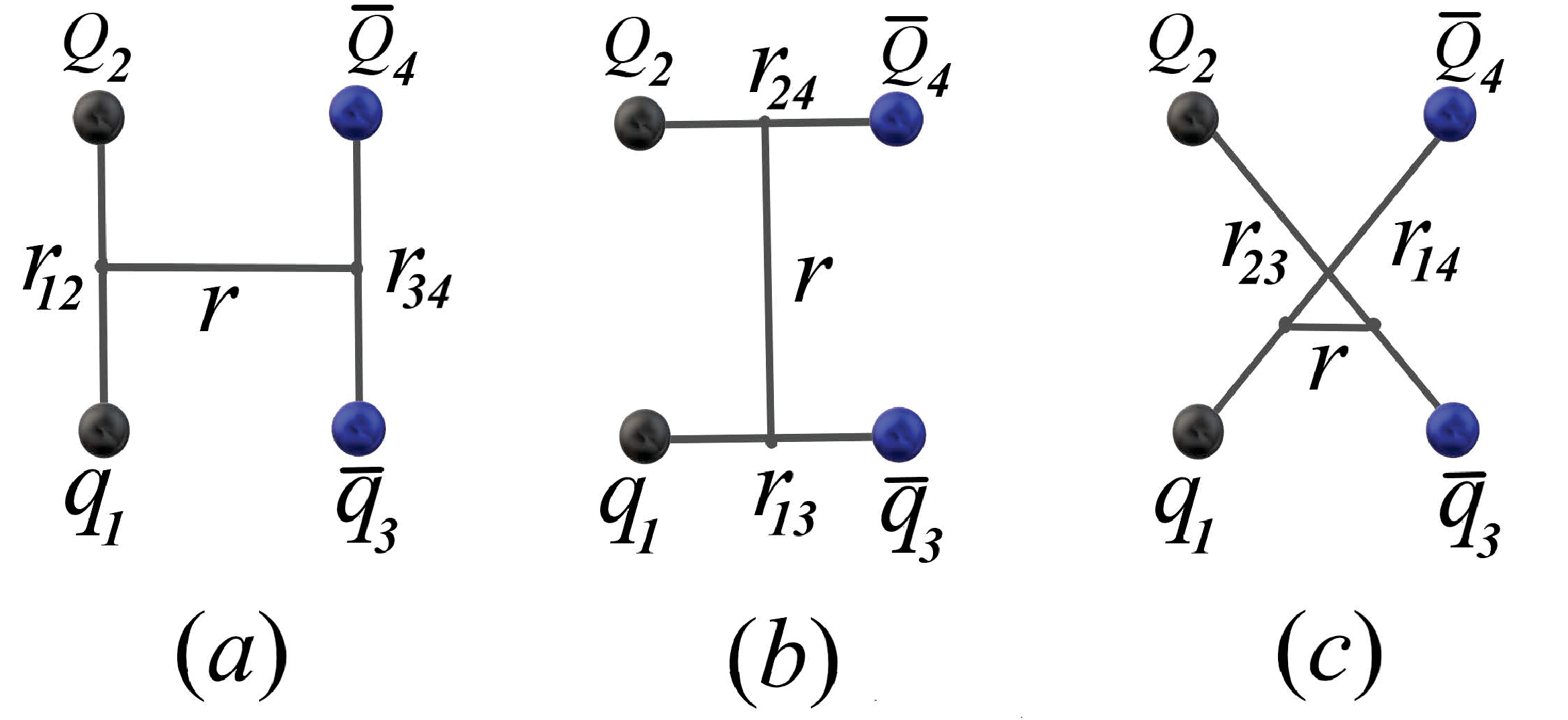}
\caption{Jacobi coordinates for a hidden-charm tetraquark state. $Q$ denotes charm quark and $q$ represents up, down or strange quark.}
\label{Jacobi}
\end{figure}

The internal motions of the quarks in a four-body system can be expressed by Jacobi coordinates shown in Fig. \ref{Jacobi}. As for Jacobi coordinates in Fig. \ref{Jacobi}(a), they can be defined as,
\begin{eqnarray}
& \boldsymbol{r}_{12}=\textbf{\emph{r}}_{2}-\textbf{\emph{r}}_{1}& \\
& \boldsymbol{r}_{34}=\textbf{\emph{r}}_{4}-\textbf{\emph{r}}_{3} & \\
& \boldsymbol{r}=\frac{m_{4}\textbf{\emph{r}}_{4}+m_{3}\textbf{\emph{r}}_{3}}{m_{3}+m_{4}}-\frac{m_{1}\textbf{\emph{r}}_{1}+m_{2}\textbf{\emph{r}}_{2}}{m_{1}+m_{2}}& \\
& \boldsymbol{R}=\frac{m_{1}\textbf{\emph{r}}_{1}+m_{2}\textbf{\emph{r}}_{2}+m_{3}\textbf{\emph{r}}_{3}+m_{4}\textbf{\emph{r}}_{4}}{m_{1}+m_{2}+m_{3}+m_{4}}
\end{eqnarray}
The other two sets of coordinates can easily be expressed by $\textbf{\emph{r}}_{12}$, $\textbf{\emph{r}}_{34}$, and $\textbf{\emph{r}}$.

The wave function of a tetraquark state is composed of color, flavor, spin, and spatial parts. In this work, the Gaussian function is employed to construct the spatial wave function of a tetrquark state, and the S-wave Gaussian function reads
\begin{eqnarray}
\phi_{n}(\boldsymbol{r})=\frac{2^{\frac{7}{4}}\nu_{n}^{\frac{3}{4}}}{\pi^\frac{1}{4}}e^{-\nu_{n}r^{2}}Y_{00}(\hat{\boldsymbol{r}})
\end{eqnarray}
with
\begin{eqnarray}\label{eq:18}
\nu_{n}=\frac{1}{r_{n}^{2}}, \quad r_{n}=r_{a1}\Big[\frac{r_{n_{\mathrm{max}}}}{r_{a 1}}\Big]^{\frac{n-1}{n_{\mathrm{max}}-1}}
\end{eqnarray}
In Eq. (\ref{eq:18}), $n_{max}$ is the maximum number of the Gaussian basis functions. It is indicated by our previous studies that $n_{max}=8$ can ensure the convergence and stability of the final results \cite{Yu:2022lak}.
Two kinds of tetraquak configurations are expressed in Fig. \ref{Jacobi}, where Fig. \ref{Jacobi}(a) is called the diquark-antidiquark configuration and the other two are meson-meson configuration. The calculations in this work are based on the Jacobi coordinates in Fig. \ref{Jacobi}(a).
Under this picture, the colorless wave function can be expressed as $|(qc)_{\bar{3}}(\bar{q}\bar{c})_{3}\rangle$ and $|(qc)_{6}(\bar{q}\bar{c})_{\bar{6}}\rangle$ which are antisymmetric and symmetric under the exchange of quarks in $qc$ or $\bar{q}\bar{c}$. The spin wave functions can be antisymmetric singlet or symmetric triplet, which can be expressed as $[qc]_{0}$, $[\bar{q}\bar{c}]_{0}$ and $[qc]_{1}$, $[\bar{q}\bar{c}]_{1}$, respectively. All possible configurations for the $qc\bar{q}\bar{c}$, $sc\bar{s}\bar{c}$ and $qc\bar{s}\bar{c}$($q=u/d$) systems are listed in Tables \ref{qcqc1}, \ref{scsc1} and \ref{qcsc1}, respectively.

The total wave function of the spin and spatial parts with angular momentum ($J$,$M$) can be written as,
\begin{equation}
\psi_{JM}=\sum_{\kappa}C_{\kappa}\Big[[(\phi_{n_{12}}(\mathbf{r}_{12})\chi_{s_{12}})_{j_{a}}(\phi_{n_{34}}(\mathbf{r}_{34})\chi_{s_{34}})_{j_{b}}]_{j}\phi_{n}(\mathbf{r})\Big]_{JM}
\end{equation}
where $C_{\kappa}$ is expansion coefficients, and $\kappa$ is the quantum numbers $\{$$n_{12}$,$s_{12}$$\cdots$$j$$\}$.

After all of the matrix elements are evaluated, the mass spectra can be obtained by solving the generalized eigenvalue problem,
\begin{flalign}
\sum_{j=1}^{n_{\mathrm{max}}^{3}}\Big(H_{ij}-E\widetilde{N}_{ij}\Big)C_{j}=0, \quad (i=1-n_{\mathrm{max}}^{3})
\end{flalign}
Here, $H_{ij}$ denotes the matrix element in the total color-flavor-spin-spatial basis, $E$ is the eigenvalue, $C_{j}$ stands for the corresponding eigenvector, and $\widetilde{N}_{ij}$ is the overlap matrix elements of the Gaussian functions, which arises from the nonorthogonality of the bases and can be expressed as,
\begin{eqnarray}
\notag
&&\widetilde{N}_{ij}\equiv \langle\phi_{n_{12}}|\phi_{n_{12}^{\prime}}\rangle\times
\langle\phi_{n_{34}}
|\phi_{n_{34}^{\prime}}\rangle\times\langle\phi_{n}|
\phi_{n^{\prime}}\rangle  \\
&&=\Big(\frac{2\sqrt{\nu_{n_{12}}\nu_{n_{12}^{\prime}}}}{\nu_{n_{12}}+\nu_{n_{12}^{\prime}}}\Big)^{\frac{3}{2}}\times
\Big(\frac{2\sqrt{\nu_{n_{34}}\nu_{n_{34}^{\prime}}}}{\nu_{n_{34}}+\nu_{n_{34}^{\prime}}}\Big)^{\frac{3}{2}}\times\Big(\frac{2\sqrt{\nu_{n}\nu_{n^{\prime}}}}{\nu_{n}+\nu_{n^{\prime}}}\Big)^{\frac{3}{2}}
\end{eqnarray}

\section{Numerical results}\label{sec3}

\begin{table}[h]
\begin{ruledtabular}\caption{Input parameters of the relativized quark model}
\label{parameters}
\begin{tabular}{c c c c c }
$m_{u/d}$(GeV)& $m_{s}$(GeV)&$m_{c}$(GeV)&$\alpha_{1}$ & $\alpha_{2}$ \\
0.22&0.419& $1.628$&  $0.25$ & $0.15$  \\ \hline
$\alpha_{3}$ &$\gamma_{1}$(GeV)&$\gamma_{2}$(GeV)& $\gamma_{3}$(GeV) &$b$(GeV$^{2}$) \\
$0.20$ &$\frac{1}{2}$&$\sqrt{10}/2$& $\sqrt{1000}/2$ & $0.23$  \\ \hline
$c$(GeV)&$\sigma_{0}$(GeV)&$s$&$\epsilon_{\mathrm{c}}$ &  $\epsilon_{\mathrm{t}}$  \\
$-0.530$ &$1.8$& $1.55$ & $-0.168$ & $0.025$  \\
\end{tabular}
\end{ruledtabular}
\end{table}
The results of the relativized quark model depend on input parameters such as the constituent quark masses and the parameters in the Hamiltonian. In most cases, the values of these parameters were determined by fitting them to the experimental data of the mesons or baryons \cite{Godfrey:1985xj,Capstick:1986ter}. As for a tetraquark state which belongs to four-body system, the parameters in the Hamiltonian should also be adjusted again by fitting reliable experimental data. Among the charmonium-like states, the observation of Z$_{c}$(3900) state invoked a large number of research works from theoretical and experimental physicists. As for its structure, one popular explanation is that it is a good candidate for diquark-antidiquark tetraquark state with quantum number to be $J^{PC}=1^{+-}$. Basing on this idea, we determine the parameters in the Hamiltonian by reproducing the measured mass of Z$_{c}$(3900) which is treated as a compact tetraquark state with $J^{PC}=1^{+-}$. All of the input parameters used in this work are listed in Table \ref{parameters} where parameters $b$ and $c$ are fixed to be $0.23$ and $-0.530$ for tetraquark system.

It is noted that a physical state of tetraquark is the mixtures of different configurations with the same quantum numbers of $J^{PC}$. Thus, the calculations are carried out in two stages. In the first stage, the masses of different configurations shown in Tables \ref{qcqc1}, \ref{scsc1} and \ref{qcsc1} are obtained by solving the Schr$\ddot{\mathrm{o}}$dinger equation. The masses and the r.m.s. radii for the ground and the first radially excited states are also presented in these tables. In the second stage, the mixing effect is considered and the masses of physical states are obtained by diagonalizing the mass matrix in the bases of eigenstates obtained in the first stage. The mass matrices, eigenvalues, and eigenvectors for the ground states and the first radial excitations are all summarized in Tables \ref{qcqcA1}$\sim$\ref{qcscA2} in Appendix.
\begin{figure}[htbp]
  \centering
   \subfigure[]{
   \begin{minipage}{5.5cm}
   \centering
   \includegraphics[width=6cm]{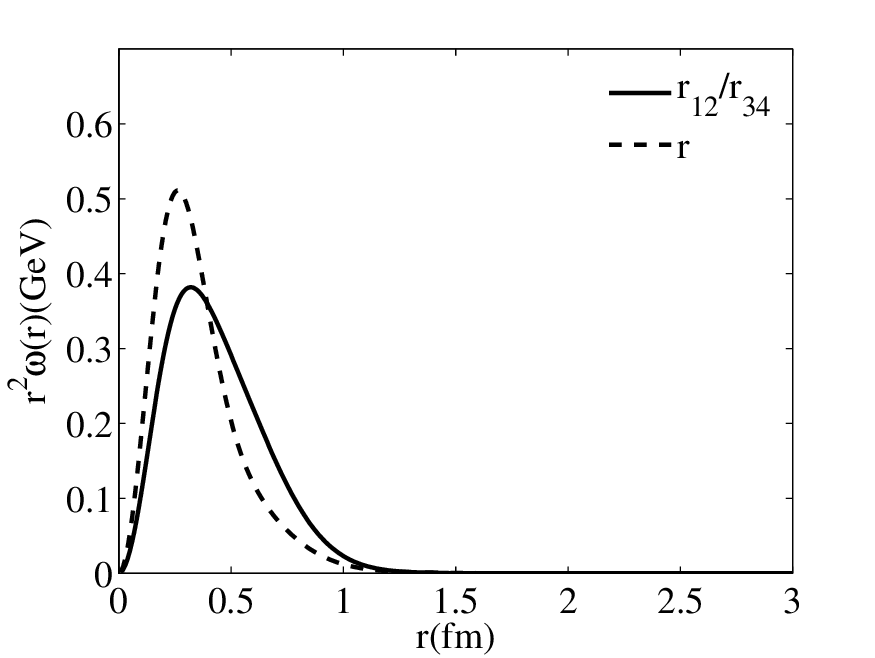}
  \end{minipage}
  }
 \subfigure[]{
   \begin{minipage}{5.5cm}
   \centering
   \includegraphics[width=6cm]{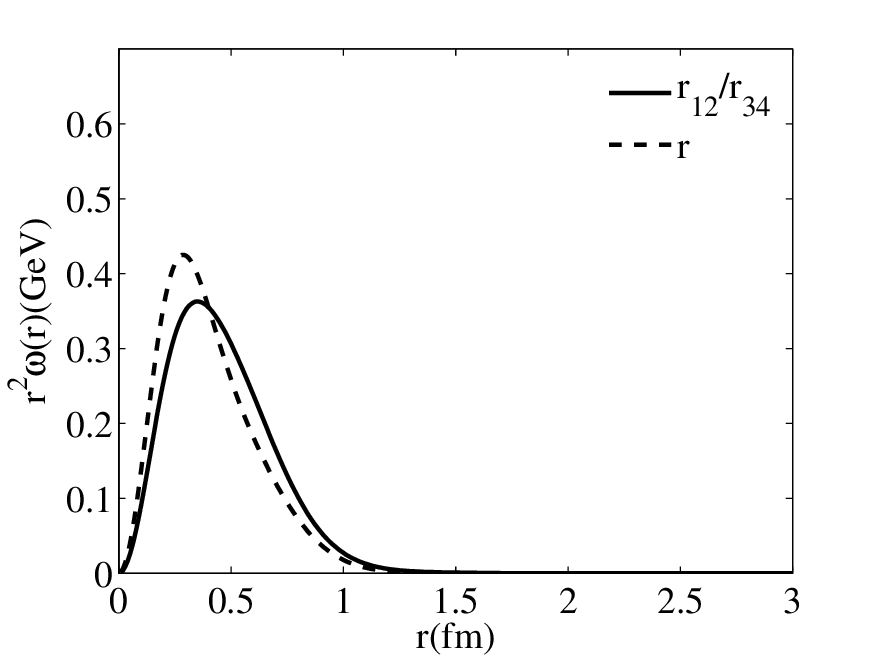}
  \end{minipage}
  }
  \caption{Radial density distributions of the ground state (a) and the first radially excited state (b) of $|(nc)_{0}^{\bar{3}}(\bar{nc})_{0}^{3}\rangle_{0}$}
\label{density1}
\end{figure}
\begin{figure}[htbp]
  \centering
   \subfigure[]{
   \begin{minipage}{5.5cm}
   \centering
   \includegraphics[width=6cm]{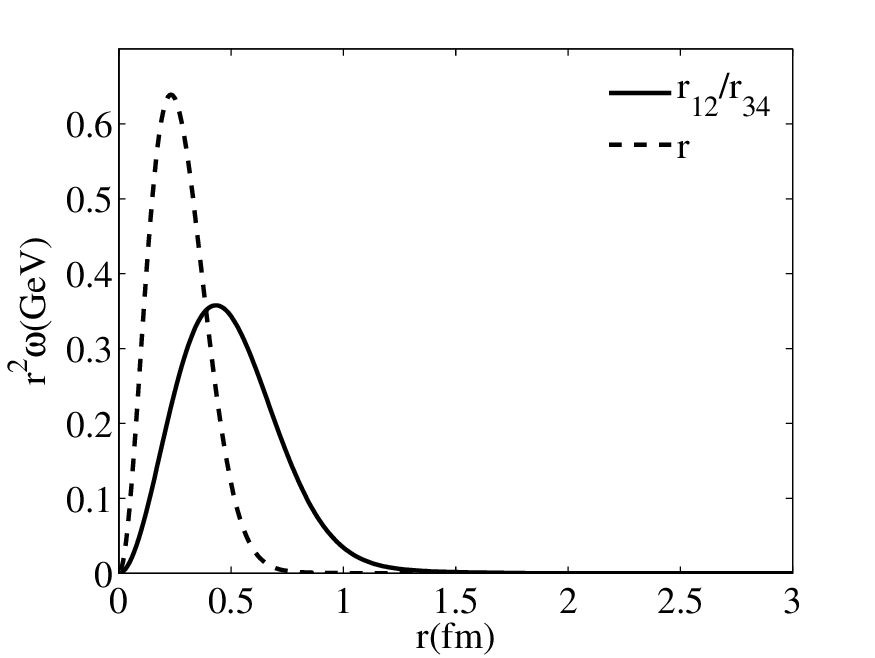}
  \end{minipage}
  }
 \subfigure[]{
   \begin{minipage}{5.5cm}
   \centering
   \includegraphics[width=6cm]{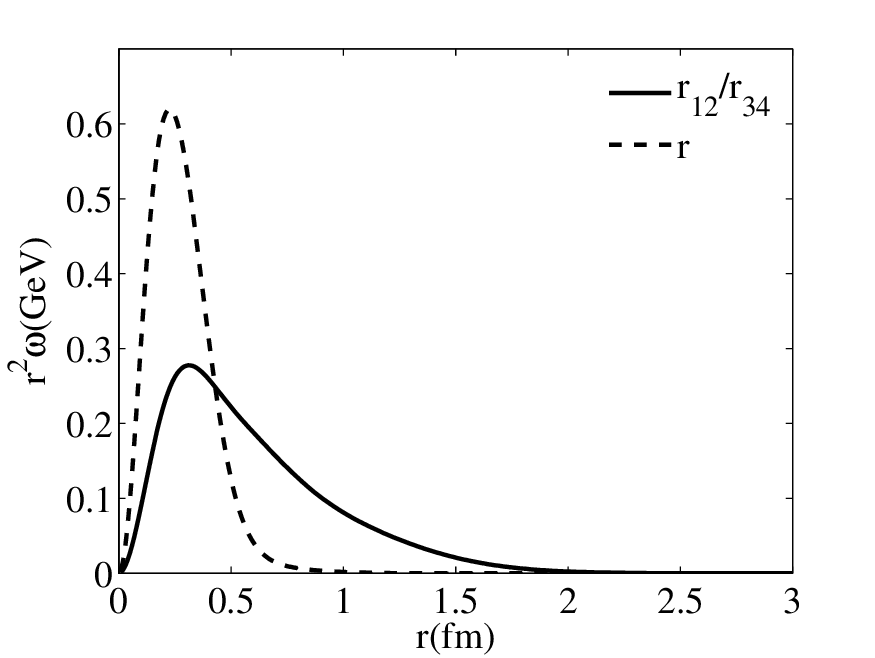}
  \end{minipage}
  }
  \caption{Radial density distributions of the ground state (a) and the first radially excited state (b) of $|(nc)_{0}^{6}(\bar{nc})_{0}^{\bar{6}}\rangle_{0}$}
\label{density2}
\end{figure}

In the one-gluon exchange(OGE) model, the interaction between the two quarks within a color-sextet diquark is repulsive, while that in the color-antitriplet one is attractive. On the other hand, the interaction between the diquark and antidiquark of $|(qc)_{6}(\bar{q}\bar{c})_{\bar{6}}\rangle$ configuration is attractive and is stronger than that of $|(qc)_{\bar{3}}(\bar{q}\bar{c})_{3}\rangle$. The competition of these two interactions commonly result in more higher energy for the $|(qc)_{6}(\bar{q}\bar{c})_{\bar{6}}\rangle$ configuration. From Table \ref{qcqc1}, one can see that $|(qc)^{6}_{0}(\bar{q}\bar{c})^{\bar{6}}_{0}\rangle_{0}$ configuration is located higher than $|(qc)_{0}^{\bar{3}}(\bar{q}\bar{c})_{0}^{3}\rangle_{0}$. Certainly, if the attractive interaction between two clusters in $|(qc)_{6}(\bar{q}\bar{c})_{\bar{6}}\rangle$ configuration is strong enough, its energy will become lower than that of $|(qc)_{\bar{3}}(\bar{q}\bar{c})_{3}\rangle$.

It is shown by Table \ref{qcqc1} that the r.m.s. radii $\sqrt{\langle r_{12/34}^{2}\rangle}$ of $|(qc)_{\bar{3}}(\bar{q}\bar{c})_{3}\rangle$ configuration are smaller than those of $|(qc)_{6}(\bar{q}\bar{c})_{\bar{6}}\rangle$. This is due to the attractive interactions between the two quarks within a color-antitriplet diquark and repulsive ones in a color-sextet diquark. On the other hand, the stronger attraction between diquark and antidiquark in $|(qc)_{6}(\bar{q}\bar{c})_{\bar{6}}\rangle$ configuration makes the situation of $\sqrt{\langle r^{2}\rangle}$ being opposite to $\sqrt{\langle r_{12/34}^{2}\rangle}$.
For example, $\sqrt{\langle r_{12/34}^{2}\rangle}$ and $\sqrt{\langle r^{2}\rangle}$ are 0.450 fm and 0.370 fm for the ground state of $|(qc)_{0}^{\bar{3}}(\bar{q}\bar{c})_{0}^{3}\rangle_{0}$ configuration, while the results are 0.554 fm and 0.273 fm for $|(qc)_{0}^{6}(\bar{q}\bar{c})_{0}^{\bar{6}}\rangle_{0}$.

To further understand the inner structures of different configurations, we also plot the radial density distributions which are obtained by the wave functions from quark model. The distribution functions are defined as,
\begin{eqnarray}
\notag
&\omega(r_{12/34})=\int|\Psi(\textbf{r}_{12},\textbf{r}_{34},\textbf{r})|^{2}d\textbf{r}d\textbf{r}_{34/12}d\Omega_{12/34} \\
&\omega(r)=\int|\Psi(\textbf{r}_{12},\textbf{r}_{34},\textbf{r})|^{2}d\textbf{r}_{12}d\textbf{r}_{34}d\Omega
\end{eqnarray}
where $\Omega_{12/34}$ and $\Omega$ are the solid angles spanned by vectors $\textbf{r}_{12/34}$ and $\textbf{r}$, respectively. For example, the density distributions of $|(qc)_{0}^{\bar{3}}(\bar{q}\bar{c})_{0}^{3}\rangle_{0}$ and $|(qc)_{0}^{6}(\bar{q}\bar{c})_{0}^{\bar{6}}\rangle_{0}$ configurations are are explicitly shown in Figs. \ref{density1} and \ref{density2}.
Firstly, it can be seen that the radial density distributions are located in the range of 1 fm, which indicates that two charmed and two light quarks are confined into a compact state. Secondly, the peaks for the first radial excitations are located more outward comparing with their ground states. Finally, by comparing Fig. \ref{density1}(a) to Fig. \ref{density2}(a), we can see that the $r^{2}_{12/34}\omega(r_{12/34})$ peak(solid line) of $|(qc)_{0}^{6}(\bar{q}\bar{c})_{0}^{\bar{6}}\rangle_{0}$ configuration is located more outward than that of $|(qc)_{0}^{\bar{3}}(\bar{q}\bar{c})_{0}^{3}\rangle_{0}$, while the situation is opposite for the peak of $r^{2}\omega(r)$(dashed line). This feature is consistent well with the characteristic of the r.m.s. radius which we have discussed in detail.
It is noted that all of these discussions are applicable to the ground state of the $qc\bar{q}\bar{c}$, $sc\bar{s}\bar{c}$ and $qc\bar{s}\bar{c}$ systems. If the radial excitation is considered, the situation will become different and complicated.
 \begin{table*}[htbp]
\begin{ruledtabular}\caption{Predicted masses (MeV) and the r.m.s.
radius (fm) of different configurations of the $qc\bar{q}\bar{c}$ ($q=u/d$) system, where I and II denote the ground and the first radially excited states, respectively.}
\label{qcqc1}
\begin{tabular}{c| c| c c c |c c c}
 \multirow{2}{*}{$J^{PC}$ }& \multirow{2}{*}{Configuration}& \multicolumn{3}{c|}{I} & \multicolumn{3}{c}{II} \\ \cline{3-8}
 \multirow{2}{*}{ }        & \multirow{2}{*}{}              & M & $\sqrt{\langle r_{12/34}^{2}\rangle}$ & $\sqrt{\langle r^{2}\rangle}$ & M & $\sqrt{\langle r_{12/34}^{2}\rangle}$ & $\sqrt{\langle r^{2}\rangle}$ \\ \hline
 \multirow{4}{*}{$0^{++}$}& $|(qc)_{1}^{\bar{3}}(\bar{q}\bar{c})_{1}^{3}\rangle_{0}$&4038 & 0.470&0.359 & 4559 & 0.540& 0.598 \\
 \multirow{4}{*}{}        & $|(qc)_{0}^{\bar{3}}(\bar{q}\bar{c})_{0}^{3}\rangle_{0}$ & 3970 & 0.450 & 0.370& 4482& 0.499& 0.621 \\
 \multirow{4}{*}{}        & $|(qc)_{1}^{6}(\bar{q}\bar{c})_{1}^{\bar{6}}\rangle_{0}$ &3834 & 0.479&0.247 & 4425 & 0.686& 0.288 \\
 \multirow{4}{*}{}        & $|(qc)_{0}^{6}(\bar{q}\bar{c})_{0}^{\bar{6}}\rangle_{0}$ & 4102 & 0.554& 0.273 &4635 &0.770 & 0.302 \\  \hline
\multirow{2}{*}{$1^{++}$}& $\frac{1}{\sqrt{2}}$[$|(qc)_{1}^{\bar{3}}(\bar{q}\bar{c})_{0}^{3}\rangle_{1}$+$|(qc)_{0}^{\bar{3}}(\bar{q}\bar{c})_{1}^{3}\rangle_{1}$] & 4039 & 0.467 &0.374  & 4546 & 0.519 &0.620  \\
\multirow{2}{*}{ }&$\frac{1}{\sqrt{2}}$[$|(qc)_{1}^{6}(\bar{q}\bar{c})_{0}^{\bar{6}}\rangle_{1}$+$|(qc)_{0}^{6}(\bar{q}\bar{c})_{1}^{\bar{6}}\rangle_{1}$]  & 4077 &0.545 &  0.289& 4615 &0.760 &  0.480  \\  \hline
 \multirow{4}{*}{$1^{+-}$}& $\frac{1}{\sqrt{2}}$[$|(qc)_{1}^{\bar{3}}(\bar{q}\bar{c})_{0}^{3}\rangle_{1}$-$|(qc)_{0}^{\bar{3}}(\bar{q}\bar{c})_{1}^{3}\rangle_{1}$] &4039 & 0.467 &0.374  & 4546 & 0.519 &0.620 \\
\multirow{4}{*}{}        & $\frac{1}{\sqrt{2}}$[$|(qc)_{1}^{6}(\bar{q}\bar{c})_{0}^{\bar{6}}\rangle_{1}$-$|(qc)_{0}^{6}(\bar{q}\bar{c})_{1}^{\bar{6}}\rangle_{1}$] & 4077 &0.545 & 0.289 & 4615 &0.760 &  0.480  \\
 \multirow{4}{*}{}        & $|(qc)_{1}^{\bar{3}}(\bar{q}\bar{c})_{1}^{3}\rangle_{1}$ &4076 & 0.477&0.447 & 4579 & 0.539& 0.394 \\
 \multirow{4}{*}{}        & $|(qc)_{1}^{6}(\bar{q}\bar{c})_{1}^{\bar{6}}\rangle_{1}$ & 4066 & 0.540& 0.288 &4606 &0.755 & 0.478 \\  \hline
\multirow{2}{*}{$2^{++}$}& $|(qc)_{1}^{\bar{3}}(\bar{q}\bar{c})_{1}^{3}\rangle_{2}$  & 4142 & 0.490 &0.389  & 4633 & 0.535 &0.631  \\
\multirow{2}{*}{ }& $|(qc)_{1}^{6}(\bar{q}\bar{c})_{1}^{\bar{6}}\rangle_{2}$ & 4147 &0.563 &0.280  & 4670 &0.783 &0.307    \\
\end{tabular}
\end{ruledtabular}
\end{table*}
\begin{table*}[htbp]
\begin{ruledtabular}\caption{The numerical results of the ground states of $qc\bar{q}\bar{c}$ ($q=u/d$) system after considering the mixture of different configurations.}
\label{qcqc2}
\begin{tabular}{c |c| c| c| c c| c c }
 \multirow{2}{*}{$J^{PC}$ } &  \multirow{2}{*}{Configuration}&\multicolumn{3}{c}{Configuration mixing (I)}   &   & \multirow{2}{*}{$1_{c}\otimes1_{c}$($\%$) }&\multirow{2}{*}{$8_{c}\otimes8_{c}$($\%$) } \\ \cline{3-6}
 \multirow{2}{*}{ }         &     \multirow{2}{*}{}                 & Eigenvalues & Mixing coefficients($\%$)&$\sqrt{\langle r_{12/34}^{2}\rangle}$ &  $\sqrt{\langle r^{2}\rangle}$ & &   \\ \hline
\multirow{4}{*}{$0^{++}$}& $|(qc)_{1}^{\bar{3}}(\bar{q}\bar{c})_{1}^{3}\rangle_{0}$ &4217    &    (36.6,1.1,0.6,61.6)& 0.523 & 0.308& 54.1&45.9\\
 \multirow{4}{*}{}& $|(qc)_{0}^{\bar{3}}(\bar{q}\bar{c})_{0}^{3}\rangle_{0}$ &   4085  &  (7.4,59.0,33.4,0.3) & 0.462& 0.333 &44.6&55.4\\
 \multirow{4}{*}{}& $|(qc)_{1}^{6}(\bar{q}\bar{c})_{1}^{\bar{6}}\rangle_{0}$ &           3947   & (49.0,15.6,2.1,33.3) & 0.497 &0.333 & 45.1& 54.9\\
 \multirow{4}{*}{}& $|(qc)_{0}^{6}(\bar{q}\bar{c})_{0}^{\bar{6}}\rangle_{0}$  &      3695  & (7.0,	24.4,63.8,4.80) & 0.475&0.291 & 56.2&43.8\\ \hline
\multirow{2}{*}{$1^{++}$}&$\frac{1}{\sqrt{2}}$[$|(qc)_{1}^{\bar{3}}(\bar{q}\bar{c})_{0}^{3}\rangle_{1}$+$|(qc)_{0}^{\bar{3}}(\bar{q}\bar{c})_{1}^{3}\rangle_{1}$] &4142    &  (38.7,61.3)  & 0.516& 0.325 & 53.8 & 46.2  \\
\multirow{2}{*}{ }& $\frac{1}{\sqrt{2}}$[$|(qc)_{1}^{6}(\bar{q}\bar{c})_{0}^{\bar{6}}\rangle_{1}$+$|(qc)_{0}^{6}(\bar{q}\bar{c})_{1}^{\bar{6}}\rangle_{1}$] &     3974             &   (61.3,38.7) & 0.499&0.344 &46.2 &53.8 \\  \hline
\multirow{4}{*}{$1^{+-}$}& $\frac{1}{\sqrt{2}}$[$|(qc)_{1}^{\bar{3}}(\bar{q}\bar{c})_{0}^{3}\rangle_{1}$-$|(qc)_{0}^{\bar{3}}(\bar{q}\bar{c})_{1}^{3}\rangle_{1}$] &         4155            &    (18.2,64.8,16.5,0.5)& 0.521 & 0.336& 39.0&61.0\\
 \multirow{4}{*}{}& $\frac{1}{\sqrt{2}}$[$|(qc)_{1}^{6}(\bar{q}\bar{c})_{0}^{\bar{6}}\rangle_{1}$-$|(qc)_{0}^{6}(\bar{q}\bar{c})_{1}^{\bar{6}}\rangle_{1}$] &   4113      & (28.3,2.7,0.7,68.2)  & 0.520& 0.316 &56.3&43.7\\
 \multirow{4}{*}{}& $|(qc)_{1}^{\bar{3}}(\bar{q}\bar{c})_{1}^{3}\rangle_{1}$ &4089     & (16.8,3.1,70.2,9.9) & 0.484 &0.418 & 60.1& 39.9\\
 \multirow{4}{*}{}& $|(qc)_{1}^{6}(\bar{q}\bar{c})_{1}^{\bar{6}}\rangle_{1}$ & 3902       & (36.7,29.4,12.7,21.3) & 0.508&0.345 & 44.6&55.4\\ \hline
 \multirow{2}{*}{$2^{++}$}&$|(qc)_{1}^{\bar{3}}(\bar{q}\bar{c})_{1}^{3}\rangle_{2}$ &4170    &   (45.1,54.9)  & 0.531& 0.333 & 51.6 & 48.4  \\
\multirow{2}{*}{ }& $|(qc)_{1}^{6}(\bar{q}\bar{c})_{1}^{\bar{6}}\rangle_{2}$  &    4119      &   (54.9,45.1) & 0.524&0.344 &48.4 &51.6 \\
\end{tabular}
\end{ruledtabular}
\end{table*}
\begin{table*}[htbp]
\begin{ruledtabular}\caption{Same as in TABLE III for the first radially excited states of $qc\bar{q}\bar{c}$ ($q=u/d$) system.}
\label{qcqc3}
\begin{tabular}{c |c| c| c| c c| c c }
 \multirow{2}{*}{$J^{PC}$ } &  \multirow{2}{*}{Configuration}&\multicolumn{3}{c}{Configuration mixing (II)}   &   & \multirow{2}{*}{$1_{c}\otimes1_{c}$($\%$) }&\multirow{2}{*}{$8_{c}\otimes8_{c}$($\%$) } \\ \cline{3-6}
 \multirow{2}{*}{ }         &     \multirow{2}{*}{}                 & Eigenvalues & Mixing coefficients($\%$)&$\sqrt{\langle r_{12/34}^{2}\rangle}$ &  $\sqrt{\langle r^{2}\rangle}$ & &   \\ \hline
\multirow{4}{*}{$0^{++}$}& $|(qc)_{1}^{\bar{3}}(\bar{q}\bar{c})_{1}^{3}\rangle_{0}$ &   4750   &  (36.1,0.5,0.3,63.0)& 0.694 & 0.434& 54.5&45.5\\
 \multirow{4}{*}{}& $|(qc)_{0}^{\bar{3}}(\bar{q}\bar{c})_{0}^{3}\rangle_{0}$ &4636 &  (5.2,50.8,43.4,0.4) & 0.590& 0.501 &48.0&52.0\\
 \multirow{4}{*}{}& $|(qc)_{1}^{6}(\bar{q}\bar{c})_{1}^{\bar{6}}\rangle_{0}$ &4460 & (51.4,15.1,1.3,32.3) & 0.621 &0.517 & 44.5& 55.5\\
 \multirow{4}{*}{}& $|(qc)_{0}^{6}(\bar{q}\bar{c})_{0}^{\bar{6}}\rangle_{0}$ &4255  & (7.2,	33.6,54.9,4.2) & 0.623&0.452 & 53.0&47.0\\ \hline
\multirow{2}{*}{$1^{++}$}&$\frac{1}{\sqrt{2}}$[$|(qc)_{1}^{\bar{3}}(\bar{q}\bar{c})_{0}^{3}\rangle_{1}$+$|(qc)_{0}^{\bar{3}}(\bar{q}\bar{c})_{1}^{3}\rangle_{1}$] &4677    &   (32.1,67.9)  & 0.692& 0.529 & 56.0 & 44.0  \\
\multirow{2}{*}{ }& $\frac{1}{\sqrt{2}}$[$|(qc)_{1}^{6}(\bar{q}\bar{c})_{0}^{\bar{6}}\rangle_{1}$+$|(qc)_{0}^{6}(\bar{q}\bar{c})_{1}^{\bar{6}}\rangle_{1}$] &4484    &   (67.9,32.1) & 0.607&0.579 &44.0 &56.0 \\  \hline
\multirow{4}{*}{$1^{+-}$}& $\frac{1}{\sqrt{2}}$[$|(qc)_{1}^{\bar{3}}(\bar{q}\bar{c})_{0}^{3}\rangle_{1}$-$|(qc)_{0}^{\bar{3}}(\bar{q}\bar{c})_{1}^{3}\rangle_{1}$] &4692                     &    (25.0,59.8,12.4,2.9)& 0.682 & 0.510& 38.5&61.5\\
 \multirow{4}{*}{}& $\frac{1}{\sqrt{2}}$[$|(qc)_{1}^{6}(\bar{q}\bar{c})_{0}^{\bar{6}}\rangle_{1}$-$|(qc)_{0}^{6}(\bar{q}\bar{c})_{1}^{\bar{6}}\rangle_{1}$] &4639 & (7.5,13.0,0.1,79.4)  & 0.740& 0.490 &59.8&40.2\\
 \multirow{4}{*}{}& $|(qc)_{1}^{\bar{3}}(\bar{q}\bar{c})_{1}^{3}\rangle_{1}$ &4587  & (22.8,1.3,74.1,1.8) & 0.543 &0.451 & 58.7& 41.4\\
 \multirow{4}{*}{}& $|(qc)_{1}^{6}(\bar{q}\bar{c})_{1}^{\bar{6}}\rangle_{1}$ &4428  & (44.8,26.1,13.4,15.8) & 0.633&0.538 & 43.1&56.9\\ \hline
 \multirow{2}{*}{$2^{++}$}&$|(qc)_{1}^{\bar{3}}(\bar{q}\bar{c})_{1}^{3}\rangle_{2}$ &4684    &   (21.7,78.3)  & 0.736& 0.400 & 59.5 & 40.5  \\
\multirow{2}{*}{ }& $|(qc)_{1}^{6}(\bar{q}\bar{c})_{1}^{\bar{6}}\rangle_{2}$ &4619  &   (78.3,21.7) & 0.598&0.576 &40.5 &59.5 \\
\end{tabular}
\end{ruledtabular}
\end{table*}
\begin{table*}[htbp]
\begin{ruledtabular}\caption{Predicted masses (MeV) and the r.m.s.
radius (fm) of different configurations of the $sc\bar{s}\bar{c}$ system, where I and II denote the ground and the first radially excited states, respectively.}
\label{scsc1}
\begin{tabular}{c| c| c c c |c c c}
 \multirow{2}{*}{$J^{PC}$ }& \multirow{2}{*}{Configuration}& \multicolumn{3}{c|}{I} & \multicolumn{3}{c}{II} \\ \cline{3-8}
 \multirow{2}{*}{ }        & \multirow{2}{*}{}              & M & $\sqrt{\langle r_{12/34}^{2}\rangle}$ & $\sqrt{\langle r^{2}\rangle}$ & M & $\sqrt{\langle r_{12/34}^{2}\rangle}$ & $\sqrt{\langle r^{2}\rangle}$ \\ \hline
 \multirow{4}{*}{$0^{++}$}& $|(sc)_{1}^{\bar{3}}(\bar{sc})_{1}^{3}\rangle_{0}$&4210 & 0.442&0.339 & 4719 & 0.517& 0.572 \\
 \multirow{4}{*}{}        & $|(sc)_{0}^{\bar{3}}(\bar{sc})_{0}^{3}\rangle_{0}$ & 4144 & 0.423 & 0.347& 4647& 0.480& 0.595 \\
 \multirow{4}{*}{}        & $|(sc)_{1}^{6}(\bar{sc})_{1}^{\bar{6}}\rangle_{0}$ &4031 & 0.463&0.240 & 4576 & 0.654& 0.283 \\
 \multirow{4}{*}{}        & $|(sc)_{0}^{6}(\bar{sc})_{0}^{\bar{6}}\rangle_{0}$ & 4237 & 0.526& 0.263 &4748 &0.725 & 0.296 \\  \hline
\multirow{2}{*}{$1^{++}$}& $\frac{1}{\sqrt{2}}$[$|(sc)_{1}^{\bar{3}}(\bar{sc})_{0}^{3}\rangle_{1}$+$|(sc)_{0}^{\bar{3}}(\bar{sc})_{1}^{3}\rangle_{1}$] &4204 & 0.438 &0.407  & 4702 & 0.496 &0.404 \\
\multirow{2}{*}{ }&$\frac{1}{\sqrt{2}}$[$|(sc)_{1}^{6}(\bar{sc})_{0}^{\bar{6}}\rangle_{1}$+$|(sc)_{0}^{6}(\bar{sc})_{1}^{\bar{6}}\rangle_{1}$]  & 4217 &0.519 & 0.286 & 4731 &0.717 &  0.417  \\  \hline
 \multirow{4}{*}{$1^{+-}$}& $\frac{1}{\sqrt{2}}$[$|(sc)_{1}^{\bar{3}}(\bar{sc})_{0}^{3}\rangle_{1}$-$|(sc)_{0}^{\bar{3}}(\bar{sc})_{1}^{3}\rangle_{1}$] &4204 & 0.438 &0.407  & 4702 & 0.496 &0.404 \\
\multirow{4}{*}{}        & $\frac{1}{\sqrt{2}}$[$|(sc)_{1}^{6}(\bar{sc})_{0}^{\bar{6}}\rangle_{1}$-$|(sc)_{0}^{6}(\bar{sc})_{1}^{\bar{6}}\rangle_{1}$] & 4217 &0.519 & 0.286 & 4731 &0.717 &  0.417  \\
 \multirow{4}{*}{}        & $|(sc)_{1}^{\bar{3}}(\bar{sc})_{1}^{3}\rangle_{1}$ &4236 & 0.446&0.379 & 4731 & 0.505& 0.434 \\
 \multirow{4}{*}{}        & $|(sc)_{1}^{6}(\bar{sc})_{1}^{\bar{6}}\rangle_{1}$ & 4208 & 0.515& 0.285 &4723 &0.713 & 0.415 \\  \hline
\multirow{2}{*}{$2^{++}$}& $|(sc)_{1}^{\bar{3}}(\bar{sc})_{1}^{3}\rangle_{2}$  & 4290 & 0.459 &0.363  & 4775 & 0.510 &0.604  \\
\multirow{2}{*}{ }& $|(sc)_{1}^{6}(\bar{sc})_{1}^{\bar{6}}\rangle_{2}$ & 4270 &0.533 &0.268  & 4776 &0.738 &0.300    \\
\end{tabular}
\end{ruledtabular}
\end{table*}
\begin{table*}[htbp]
\begin{ruledtabular}\caption{The numerical results of the ground states of $sc\bar{s}\bar{c}$ system after considering the mixture of different configurations.}
\label{scsc2}
\begin{tabular}{c |c| c| c| c c| c c }
 \multirow{2}{*}{$J^{PC}$ } &  \multirow{2}{*}{Configuration}&\multicolumn{3}{c}{Configuration mixing (I)}   &   & \multirow{2}{*}{$1_{c}\otimes1_{c}$($\%$) }&\multirow{2}{*}{$8_{c}\otimes8_{c}$($\%$) } \\ \cline{3-6}
 \multirow{2}{*}{ }         &     \multirow{2}{*}{}                 & Eigenvalues & Mixing coefficients($\%$)&$\sqrt{\langle r_{12/34}^{2}\rangle}$ &  $\sqrt{\langle r^{2}\rangle}$ & &   \\ \hline
\multirow{4}{*}{$0^{++}$}& $|(sc)_{1}^{\bar{3}}(\bar{sc})_{1}^{3}\rangle_{0}$ &4335      & (43.2,0.3,0.14,56.4)& 0.491 & 0.298& 52.2&47.8\\
 \multirow{4}{*}{}& $|(sc)_{0}^{\bar{3}}(\bar{sc})_{0}^{3}\rangle_{0}$ &4229  & (2.4,66.6,30.7,0.3)  & 0.436& 0.318 &43.6&56.4\\
 \multirow{4}{*}{}& $|(sc)_{1}^{6}(\bar{sc})_{1}^{\bar{6}}\rangle_{0}$ &4119  & (51.8,5.9, 1.0,41.3) & 0.478 &0.309 & 47.5& 52.6\\
 \multirow{4}{*}{}&$|(sc)_{0}^{6}(\bar{sc})_{0}^{\bar{6}}\rangle_{0}$  &3940  & (2.6,27.2,68.2,2.0) & 0.453&0.276 & 56.8&43.2\\ \hline
\multirow{2}{*}{$1^{++}$}&$\frac{1}{\sqrt{2}}$[$|(sc)_{1}^{\bar{3}}(\bar{sc})_{0}^{3}\rangle_{1}$+$|(sc)_{0}^{\bar{3}}(\bar{sc})_{1}^{3}\rangle_{1}$] &4274    &   (44.9,55.1)  & 0.485& 0.346 & 51.8 & 48.2  \\
\multirow{2}{*}{ }& $\frac{1}{\sqrt{2}}$[$|(sc)_{1}^{6}(\bar{sc})_{0}^{\bar{6}}\rangle_{1}$+$|(sc)_{0}^{6}(\bar{sc})_{1}^{\bar{6}}\rangle_{1}$] &4147                  &   (55.1,44.9) & 0.476&0.358 &48.2 &51.8 \\  \hline
\multirow{4}{*}{$1^{+-}$}& $\frac{1}{\sqrt{2}}$[$|(sc)_{1}^{\bar{3}}(\bar{sc})_{0}^{3}\rangle_{1}$-$|(sc)_{0}^{\bar{3}}(\bar{sc})_{1}^{3}\rangle_{1}$] &4276                     &    (37.0,55.8,7.1,0.2)& 0.486 & 0.342& 35.8&64.2\\
 \multirow{4}{*}{}& $\frac{1}{\sqrt{2}}$[$|(sc)_{1}^{6}(\bar{sc})_{0}^{\bar{6}}\rangle_{1}$-$|(sc)_{0}^{6}(\bar{sc})_{1}^{\bar{6}}\rangle_{1}$] &4242         & (13.7,0.1,85.6,0.6)  & 0.445& 0.382 &62.0&38.0\\
 \multirow{4}{*}{}& $|(sc)_{1}^{\bar{3}}(\bar{sc})_{1}^{3}\rangle_{1}$ &4222  & (7.5,7.8,0.1,84.6) & 0.510 &0.296 & 61.6& 38.4\\
 \multirow{4}{*}{}& $|(sc)_{1}^{6}(\bar{sc})_{1}^{\bar{6}}\rangle_{1}$ & 4124  & (41.9,36.2,7.2,14.7) & 0.481&0.348 & 40.6&59.4\\ \hline
 \multirow{2}{*}{$2^{++}$}& $|(sc)_{1}^{\bar{3}}(\bar{sc})_{1}^{3}\rangle_{2}$&4295    &   (83.7,16.3) & 0.472& 0.349 & 38.8 & 61.2  \\
\multirow{2}{*}{ }& $|(sc)_{1}^{6}(\bar{sc})_{1}^{\bar{6}}\rangle_{2}$ &4265           &   (16.3,83.7) & 0.522&0.286 &61.2 &38.8 \\
\end{tabular}
\end{ruledtabular}
\end{table*}
\begin{table*}[htbp]
\begin{ruledtabular}\caption{Same as in TABLE VI for the first radially excited states of $sc\bar{s}\bar{c}$ system.}
\label{scsc3}
\begin{tabular}{c |c| c| c| c c| c c }
 \multirow{2}{*}{$J^{PC}$ } &  \multirow{2}{*}{Configuration}&\multicolumn{3}{c}{Configuration mixing (II)}   &   & \multirow{2}{*}{$1_{c}\otimes1_{c}$($\%$) }&\multirow{2}{*}{$8_{c}\otimes8_{c}$($\%$) } \\ \cline{3-6}
 \multirow{2}{*}{ }         &     \multirow{2}{*}{}                 & Eigenvalues & Mixing coefficients($\%$)&$\sqrt{\langle r_{12/34}^{2}\rangle}$ &  $\sqrt{\langle r^{2}\rangle}$ & &   \\ \hline
\multirow{4}{*}{$0^{++}$}& $|(sc)_{1}^{\bar{3}}(\bar{sc})_{1}^{3}\rangle_{0}$ &4851      &    (43.4,0.2,0.1,56.4)& 0.643 & 0.438& 52.1&47.9\\
 \multirow{4}{*}{}& $|(sc)_{0}^{\bar{3}}(\bar{sc})_{0}^{3}\rangle_{0}$ &4753 & (1.8,56.0,38.0, 0.3)  & 0.546& 0.485 &46.1&53.9\\
 \multirow{4}{*}{}& $|(sc)_{1}^{6}(\bar{sc})_{1}^{\bar{6}}\rangle_{0}$ &4621 & (52.1,5.9,0.7,41.3) & 0.611 &0.476 & 47.4& 52.6\\
 \multirow{4}{*}{}& $|(sc)_{0}^{6}(\bar{sc})_{0}^{\bar{6}}\rangle_{0}$ &4466 & (2.7,34.2,61.3,2.0) & 0.599&0.425 & 54.5&45.7\\ \hline
\multirow{2}{*}{$1^{++}$}&$\frac{1}{\sqrt{2}}$[$|(sc)_{1}^{\bar{3}}(\bar{sc})_{0}^{3}\rangle_{1}$+$|(sc)_{0}^{\bar{3}}(\bar{sc})_{1}^{3}\rangle_{1}$] &4784    &  (42.6,57.4)  & 0.632& 0.411 & 52.4 & 47.6  \\
\multirow{2}{*}{ }& $\frac{1}{\sqrt{2}}$[$|(sc)_{1}^{6}(\bar{sc})_{0}^{\bar{6}}\rangle_{1}$+$|(sc)_{0}^{6}(\bar{sc})_{1}^{\bar{6}}\rangle_{1}$] &4641                  &  (57.4,42.6) & 0.600&0.409 &47.6 &52.4 \\  \hline
\multirow{4}{*}{$1^{+-}$}& $\frac{1}{\sqrt{2}}$[$|(sc)_{1}^{\bar{3}}(\bar{sc})_{0}^{3}\rangle_{1}$-$|(sc)_{0}^{\bar{3}}(\bar{sc})_{1}^{3}\rangle_{1}$] &4804                     &    (25.5,52.3,19.3,3.0)& 0.629 & 0.417& 40.7&59.3\\
 \multirow{4}{*}{}& $\frac{1}{\sqrt{2}}$[$|(sc)_{1}^{6}(\bar{sc})_{0}^{\bar{6}}\rangle_{1}$-$|(sc)_{0}^{6}(\bar{sc})_{1}^{\bar{6}}\rangle_{1}$] &4743         &  (6.6,11.2,32.4,49.8) & 0.640& 0.421 &60.7&39.3\\
 \multirow{4}{*}{}& $|(sc)_{1}^{\bar{3}}(\bar{sc})_{1}^{3}\rangle_{1}$ &4724 & (27.5,2.2,34.9,35.5) & 0.590 &0.414 & 56.9& 43.1\\
 \multirow{4}{*}{}& $|(sc)_{1}^{6}(\bar{sc})_{1}^{\bar{6}}\rangle_{1}$ & 4616 & (40.6,34.3,13.4,11.7) & 0.608&0.414 & 41.7&58.3\\ \hline
 \multirow{2}{*}{$2^{++}$}& $|(sc)_{1}^{\bar{3}}(\bar{sc})_{1}^{3}\rangle_{2}$&4788    &   (47.9,52.1)  & 0.639& 0.471 & 50.7 & 49.3  \\
\multirow{2}{*}{ }& $|(sc)_{1}^{6}(\bar{sc})_{1}^{\bar{6}}\rangle_{2}$ &4763  &   (52.1,47.9) & 0.630&0.483 &49.3 &50.7 \\
\end{tabular}
\end{ruledtabular}
\end{table*}
\begin{table*}[htbp]
\begin{ruledtabular}\caption{Predicted masses (MeV) and the r.m.s.
radius (fm) of different configurations of the $qc\bar{s}\bar{c}$ ($q=u/d$) system, where I and II denote the ground and the first radially excited states, respectively.}
\label{qcsc1}
\begin{tabular}{c| c| c c c c|c c c c}
 \multirow{2}{*}{$J^{P}$ }& \multirow{2}{*}{Configuration}& \multicolumn{4}{c|}{I} & \multicolumn{4}{c}{II} \\ \cline{3-10}
 \multirow{2}{*}{ }        & \multirow{2}{*}{}              & M & $\sqrt{\langle r_{12}^{2}\rangle}$ & $\sqrt{\langle r_{34}^{2}\rangle}$& $\sqrt{\langle r^{2}\rangle}$ & M & $\sqrt{\langle r_{12}^{2}\rangle}$ & $\sqrt{\langle r_{34}^{2}\rangle}$& $\sqrt{\langle r^{2}\rangle}$ \\ \hline
 \multirow{4}{*}{$0^{+}$}& $|(qc)_{1}^{\bar{3}}(\bar{sc})_{1}^{3}\rangle_{0}$&4129 & 0.466&0.447 &0.350 & 4643 & 0.533& 0.524&0.586 \\
 \multirow{4}{*}{}        & $|(qc)_{0}^{\bar{3}}(\bar{sc})_{0}^{3}\rangle_{0}$ & 4059 & 0.444 & 0.430& 0.359& 4567& 0.494& 0.486 & 0.608\\
 \multirow{4}{*}{}        & $|(qc)_{1}^{6}(\bar{sc})_{1}^{\bar{6}}\rangle_{0}$ &3946 & 0.477&0.470 & 0.245&4512 & 0.660& 0.683& 0.287\\
 \multirow{4}{*}{}        & $|(qc)_{0}^{6}(\bar{sc})_{0}^{\bar{6}}\rangle_{0}$ & 4174 & 0.542& 0.537 &0.269&4697 &0.741 & 0.756& 0.301\\  \hline
\multirow{6}{*}{$1^{+}$}& $|(qc)_{1}^{\bar{3}}(\bar{sc})_{0}^{3}\rangle_{1}$ &4130 & 0.476&0.432 & 0.363& 4631 & 0.532& 0.484& 0.608\\
\multirow{6}{*}{ }&$|(qc)_{0}^{\bar{3}}(\bar{sc})_{1}^{3}\rangle_{1}$  & 4118 & 0.446 & 0.457&0.363 & 4620& 0.492& 0.521 & 0.607\\
 \multirow{6}{*}{}&$|(qc)_{1}^{6}(\bar{sc})_{0}^{\bar{6}}\rangle_{1}$ &4149 & 0.528 &0.534  & 0.287&4676 & 0.721 &0.756 &0.452\\
\multirow{6}{*}{}        & $|(qc)_{0}^{6}(\bar{sc})_{1}^{\bar{6}}\rangle_{1}$ & 4156 &0.540 & 0.525 & 0.287& 4681 &0.739 &  0.741 & 0.452\\
 \multirow{6}{*}{}        & $|(qc)_{1}^{\bar{3}}(\bar{sc})_{1}^{3}\rangle_{1}$ &4160 & 0.472&0.454 & 0.428& 4664 & 0.532& 0.522 & 0.403 \\
 \multirow{6}{*}{}        & $|(qc)_{1}^{6}(\bar{sc})_{1}^{\bar{6}}\rangle_{1}$ & 4142 & 0.529& 0.526 &0.287 &4670 &0.725 & 0.745 & 0.450\\  \hline
\multirow{2}{*}{$2^{+}$}& $|(qc)_{1}^{\bar{3}}(\bar{sc})_{1}^{3}\rangle_{2}$  & 4217 & 0.483 &0.465& 0.376 & 4705 & 0.530 &0.516&  0.617 \\
\multirow{2}{*}{ }& $|(qc)_{1}^{6}(\bar{sc})_{1}^{\bar{6}}\rangle_{2}$ & 4210 &0.547 &0.546& 0.275 & 4726 &0.752 &0.770  & 0.305 \\
\end{tabular}
\end{ruledtabular}
\end{table*}
\begin{table*}[htbp]
\begin{ruledtabular}\caption{The numerical results of the ground states of $qc\bar{s}\bar{c}$ ($q=u/d$) system after considering the mixture of different configurations.}
\label{qcsc2}
\begin{tabular}{c |c| c| c| c c c| c c }
 \multirow{2}{*}{$J^{P}$ } &  \multirow{2}{*}{Configuration}&\multicolumn{4}{c}{Configuration mixing (I)}   &   & \multirow{2}{*}{$1_{c}\otimes1_{c}$($\%$) }&\multirow{2}{*}{$8_{c}\otimes8_{c}$($\%$) } \\ \cline{3-7}
 \multirow{2}{*}{ }         &     \multirow{2}{*}{}       & Eigenvalues & Mixing coefficients($\%$)&$\sqrt{\langle r_{12}^{2}\rangle}$ & $\sqrt{\langle r_{34}^{2}\rangle}$& $\sqrt{\langle r^{2}\rangle}$ & &   \\ \hline
\multirow{4}{*}{$0^{+}$}& $|(qc)_{1}^{\bar{3}}(\bar{sc})_{1}^{3}\rangle_{0}$ &4275      &    (39.6,0.4, 0.2,59.8)& 0.513&0.503 & 0.304& 53.4&46.6\\
 \multirow{4}{*}{}& $|(qc)_{0}^{\bar{3}}(\bar{sc})_{0}^{3}\rangle_{0}$ &4155 &  (4.2,62.9,32.6,0.3) & 0.456& 0.444&0.326 &44.3&55.7\\
 \multirow{4}{*}{}& $|(qc)_{1}^{6}(\bar{sc})_{1}^{\bar{6}}\rangle_{0}$ &4039 & (52.3,9.4,1.2,37.1) & 0.494 &0.481 &0.322 & 46.1& 53.9\\
 \multirow{4}{*}{}& $|(qc)_{0}^{6}(\bar{sc})_{0}^{\bar{6}}\rangle_{0}$ &3840  & (4.1,27.1,65.9,2.9) & 0.470&0.461 &0.286 & 56.3&43.7\\ \hline
\multirow{6}{*}{$1^{+}$}&$|(qc)_{1}^{\bar{3}}(\bar{sc})_{0}^{3}\rangle_{1}$ &4251    &   (12.8,28.5,19.4,10.2,9.3,19.8)  & 0.496& 0.491 &0.335 & 49.8 & 50.2  \\
\multirow{6}{*}{ }& $|(qc)_{0}^{\bar{3}}(\bar{sc})_{1}^{3}\rangle_{1}$ &4215 &  (24.2,9.5,7.3,46.1,1.6,11.2)  & 0.513&0.497 &0.317 &54.8 &45.2 \\
\multirow{6}{*}{}& $|(qc)_{1}^{6}(\bar{sc})_{0}^{\bar{6}}\rangle_{1}$ &4182  &  (0.7,0.1,32.8,3.6,58.1,4.8)  & 0.497 & 0.488&0.376& 47.1&52.9\\
 \multirow{6}{*}{}& $|(qc)_{0}^{6}(\bar{sc})_{1}^{\bar{6}}\rangle_{1}$ &4131   & (28.7,3.3,11.4,5.9,17.2,   33.5)  & 0.503& 0.487&0.340 &50.3&49.7\\
 \multirow{6}{*}{}& $|(qc)_{1}^{\bar{3}}(\bar{sc})_{1}^{3}\rangle_{1}$ &4051  & (31.2,9.0,0.2,28.8,1.1,29.5) & 0.579 &0.565  &0.374 & 52.8& 47.2\\
 \multirow{6}{*}{}& $|(qc)_{1}^{6}(\bar{sc})_{1}^{\bar{6}}\rangle_{1}$ & 4034  & (2.3,49.7,28.9,5.2,12.7,1.2) & 0.506& 0.491 &0.341 & 45.2&54.8\\ \hline
 \multirow{2}{*}{$2^{+}$}& $|(qc)_{1}^{\bar{3}}(\bar{sc})_{1}^{3}\rangle_{2}$&4299    &   (61.3,38.7)  & 0.509&0.498   &0.340 & 46.2 & 53.8  \\
\multirow{2}{*}{ }& $|(qc)_{1}^{6}(\bar{sc})_{1}^{\bar{6}}\rangle_{2}$ &4198 &  (38.7,61.3)  & 0.523& 0.516 &0.318 &53.8 &46.2 \\
\end{tabular}
\end{ruledtabular}
\end{table*}
\begin{table*}[htbp]
\begin{ruledtabular}\caption{Same as in TABLE IX for the first radially excited states of $qc\bar{s}\bar{c}$ ($q=u/d$) system.}
\label{qcsc3}
\begin{tabular}{c |c| c| c| c c c| c c }
 \multirow{2}{*}{$J^{P}$ } &  \multirow{2}{*}{Configuration}&\multicolumn{4}{c}{Configuration mixing (II)}   &   & \multirow{2}{*}{$1_{c}\otimes1_{c}$($\%$) }&\multirow{2}{*}{$8_{c}\otimes8_{c}$($\%$) } \\ \cline{3-7}
 \multirow{2}{*}{ }         &     \multirow{2}{*}{}       & Eigenvalues & Mixing coefficients($\%$)&$\sqrt{\langle r_{12}^{2}\rangle}$ & $\sqrt{\langle r_{34}^{2}\rangle}$& $\sqrt{\langle r^{2}\rangle}$ & &   \\ \hline
\multirow{4}{*}{$0^{+}$}& $|(qc)_{1}^{\bar{3}}(\bar{sc})_{1}^{3}\rangle_{0}$ &4799      &    (38.8,0.2,0.1,60.8)& 0.667&0.775 & 0.435& 53.7&46.3\\
 \multirow{4}{*}{}& $|(qc)_{0}^{\bar{3}}(\bar{sc})_{0}^{3}\rangle_{0}$ &4692 & (3.0,54.9,41.7,0.4)  & 0.571&0.579 &0.498 &47.4&52.6\\
 \multirow{4}{*}{}& $|(qc)_{1}^{6}(\bar{sc})_{1}^{\bar{6}}\rangle_{0}$ &4548 & (54.2,8.8,0.7,36.2) & 0.614 &0.616 &0.502 & 45.6& 54.4\\
 \multirow{4}{*}{}& $|(qc)_{0}^{6}(\bar{sc})_{0}^{\bar{6}}\rangle_{0}$ &4380 & (4.0,36.1,57.5,2.5) & 0.603& 0.615&0.444 & 53.3&46.7\\ \hline
\multirow{6}{*}{$1^{+}$}&$|(qc)_{1}^{\bar{3}}(\bar{sc})_{0}^{3}\rangle_{1}$ &4788    &   (0.1,35.4,13.2,0.5,24.2,26.6)   & 0.605& 0.623 &0.502 & 46.8 & 53.2  \\
\multirow{6}{*}{ }& $|(qc)_{0}^{\bar{3}}(\bar{sc})_{1}^{3}\rangle_{1}$ &4746 &  (35.4,0.0,9.5,46.2,8.8,0.0)  & 0.653& 0.644&0.509 &51.9 &48.1 \\
\multirow{6}{*}{}& $|(qc)_{1}^{6}(\bar{sc})_{0}^{\bar{6}}\rangle_{1}$ &4698  & (0.7,1.0,45.7,22.6,30.0,0)  & 0.671 & 0.687&0.441& 56.1&43.9\\
 \multirow{6}{*}{}& $|(qc)_{0}^{6}(\bar{sc})_{1}^{\bar{6}}\rangle_{1}$ &4661  & (17.2,0.9,13.8,9.0,8.4,50.7)  & 0.680&0.689 &0.479 &57.9&42.1\\
 \multirow{6}{*}{}& $|(qc)_{1}^{\bar{3}}(\bar{sc})_{1}^{3}\rangle_{1}$ &4551  & (28.9,30.0,4.6,13.4,0.6,22.5) & 0.762 &0.775  &0.525 & 46.8& 53.2\\
 \multirow{6}{*}{}&$|(qc)_{1}^{6}(\bar{sc})_{1}^{\bar{6}}\rangle_{1}$  & 4498 & (17.7,32.6,13.2,8.4,28.0,0.1) & 0.680& 0.690 &0.480 & 40.6&59.4\\ \hline
 \multirow{2}{*}{$2^{+}$}&$|(qc)_{1}^{\bar{3}}(\bar{sc})_{1}^{3}\rangle_{2}$ &4735    &   ( 23.7,76.2)  & 0.705&0.718   &0.401 & 58.7 & 41.2  \\
\multirow{2}{*}{ }& $|(qc)_{1}^{6}(\bar{sc})_{1}^{\bar{6}}\rangle_{2}$ &4696 &  (76.2,23.7)  & 0.590& 0.586 &0.559 &41.2 &58.7 \\
\end{tabular}
\end{ruledtabular}
\end{table*}
\begin{figure*}[htbp]
\centering
\includegraphics[width=0.78\textwidth,trim=20 0 0 0,clip]{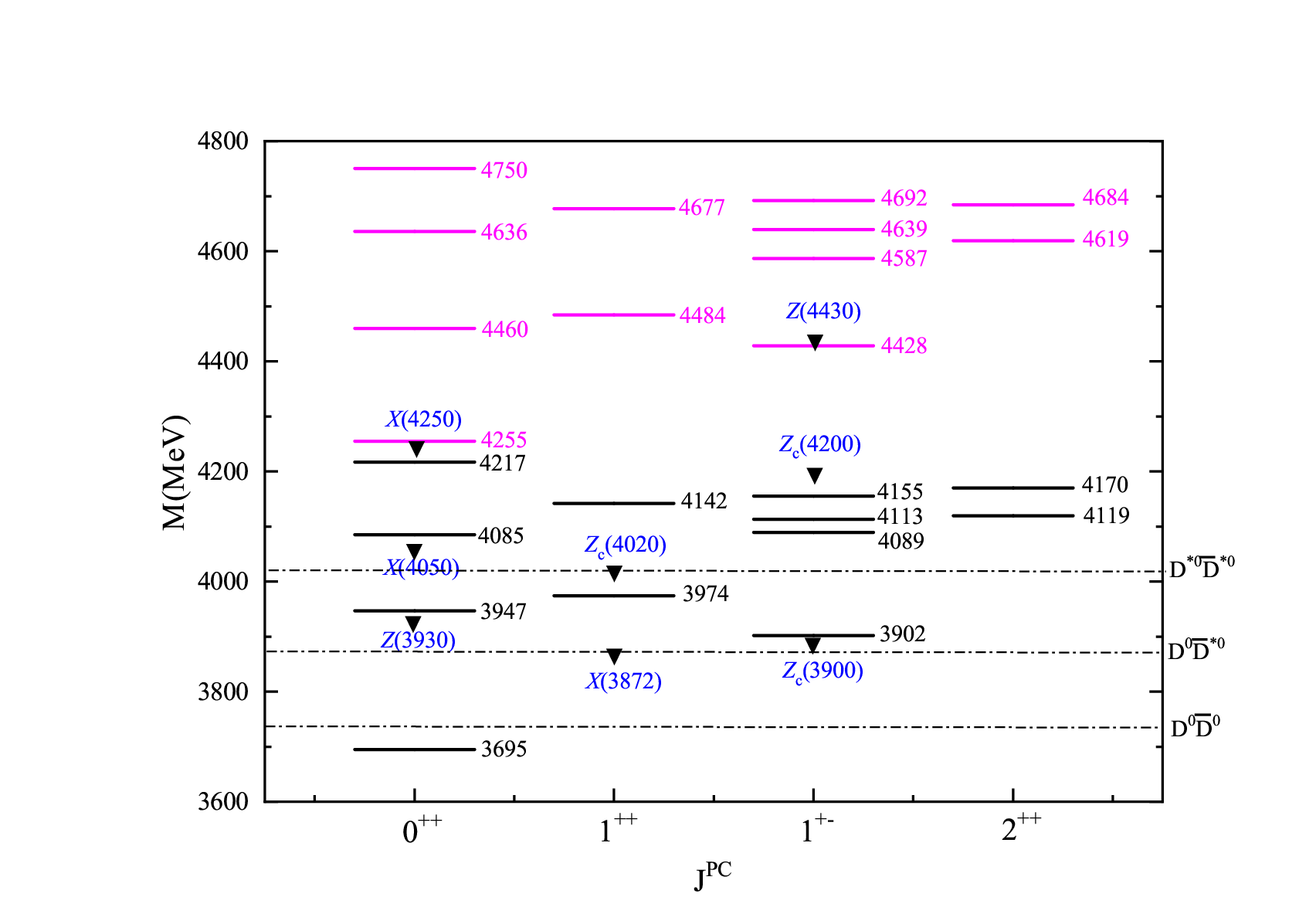}
\caption{Mass spectrum for the $qc\bar{q}\bar{c}$ ($q=u/d$) tetraquark system. Black and pink solid lines denote the ground and the first radially excited states, respectively. Inverted triangles represent experimental data, dotted lines are corresponding charm meson pair thresholds.}
\label{qcqcfig}
\end{figure*}
\begin{figure*}[htbp]
\centering
\includegraphics[width=0.8\textwidth]{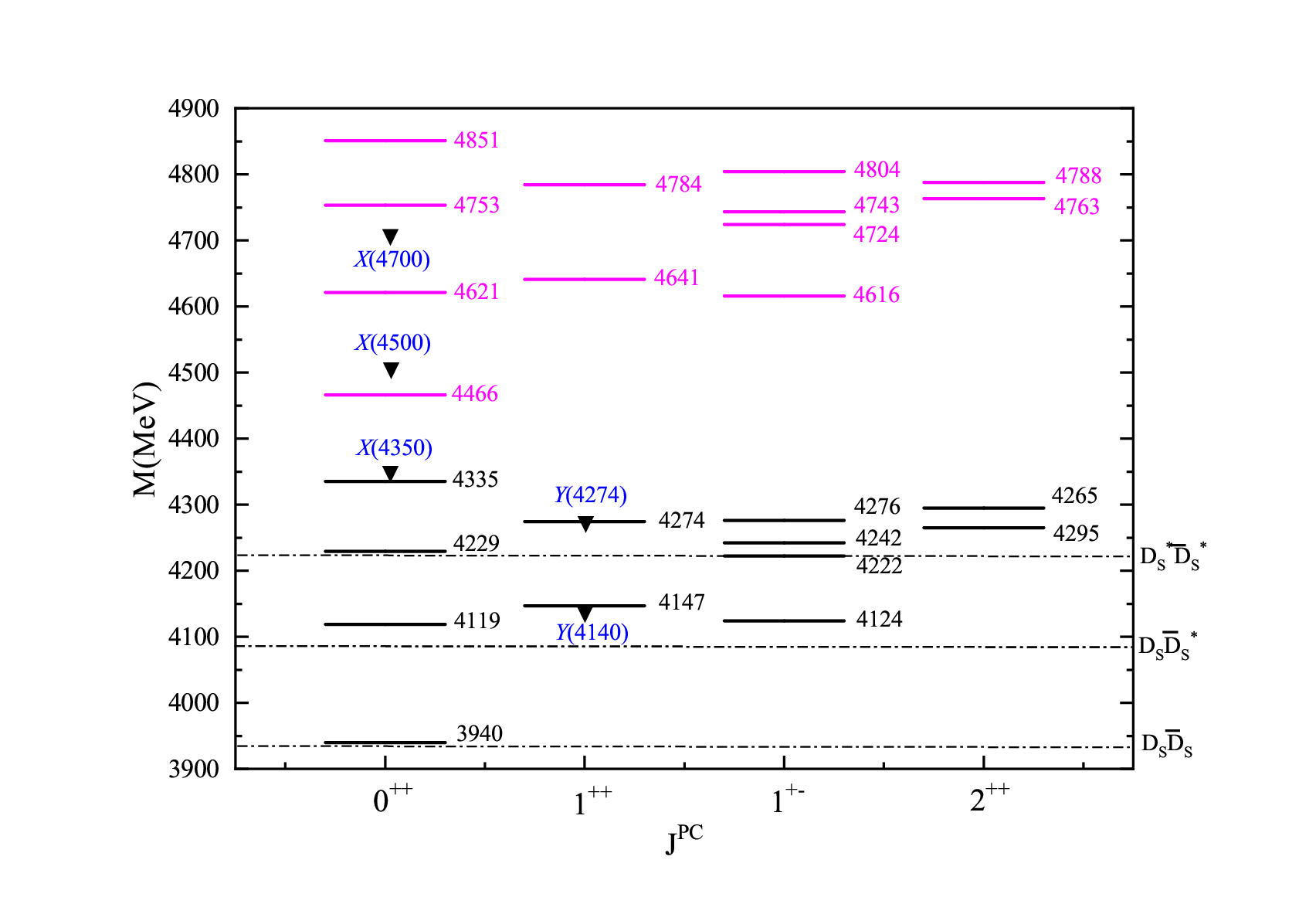}
\caption{Same as in Fig.4 for $sc\bar{s}\bar{c}$ tetraquark system}
\label{scscfig}
\end{figure*}
\begin{figure*}[htbp]
\centering
\includegraphics[width=0.8\textwidth]{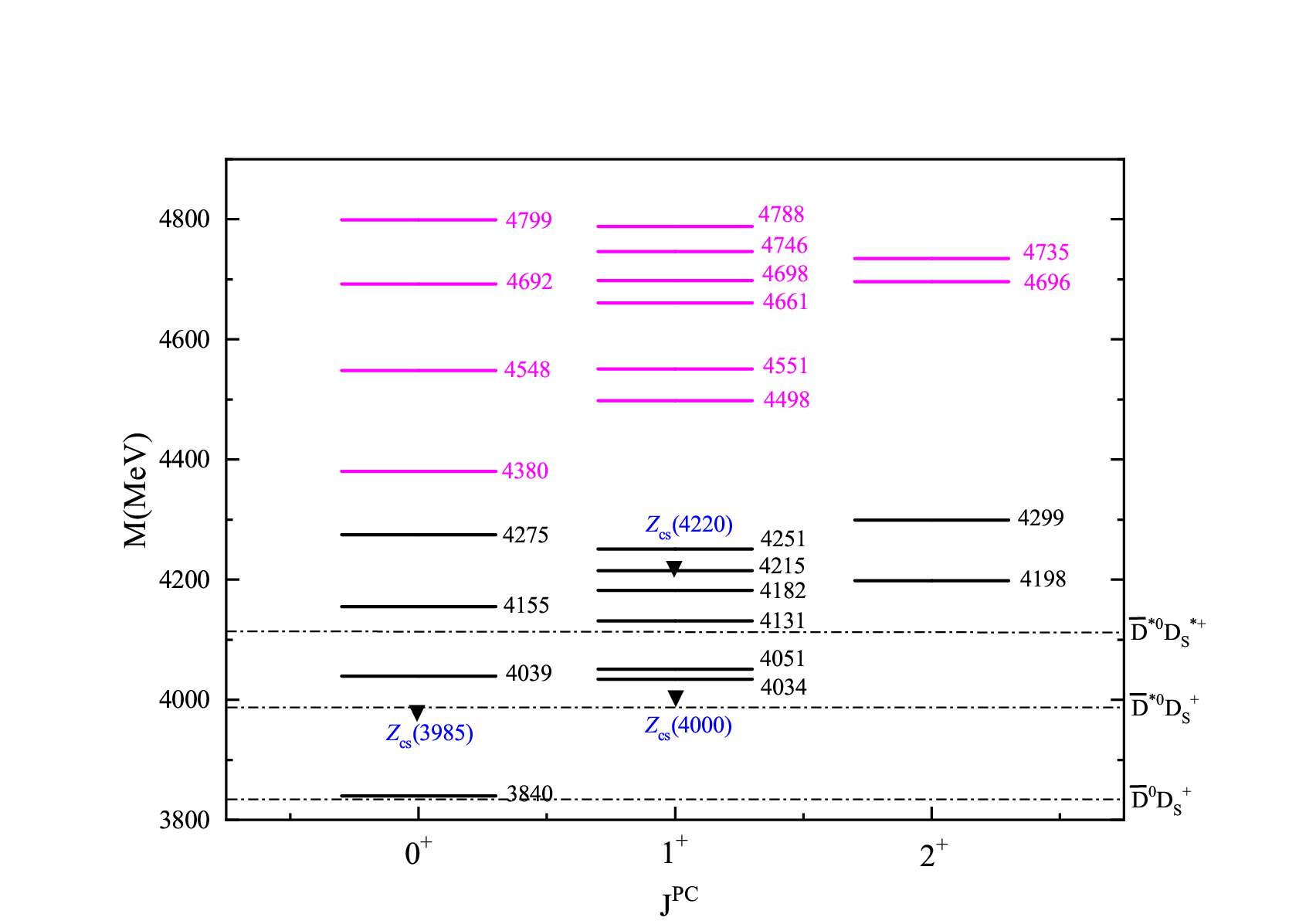}
\caption{Same as in Fig.4 for $qc\bar{s}\bar{c}$ ($q=u/d$) tetraquark system}
\label{qcscfig}
\end{figure*}

Considering mixtures of different configurations, we obtain the eigenvalues and eigenvectors by diagonalizing mass matrices. The mass matrices, eigenvalues and eigenvectors are all listed in the tables in Appendix. The percentages of different configurations on physical states are all listed in the Tables \ref{qcqc2}$\sim$\ref{qcqc3}($qc\bar{q}\bar{c}$ system), \ref{scsc2}$\sim$\ref{scsc3}($sc\bar{s}\bar{c}$ system) and \ref{qcsc2}$\sim$\ref{qcsc3}($qc\bar{s}\bar{c}$ system). It can be seen from these tables that the masses of the lowest physical states are pulled down by the mixing effect, while the highest states are raised up. Besides, one can also see that some physical states have strong mixing effect, while some states are dominated by fewer components. For the ground state of $qc\bar{q}\bar{c}$ system as an example, the percentage of the $1^{+-}$(3902) state on different configurations are (36.7$\%$, 29.4$\%$, 12.7$\%$, 21.3$\%$), while for the $1^{+-}$(4113) state its values are (28.3$\%$, 2.7$\%$, 0.7$\%$, 68.2$\%$).

Actually there are several interpretations about the hidden-charm tetraquark states such as the diquark-antidiquark configuration, the meson-meson configuration, molecule states, even mixtures of either two of them, and the coupled-channel effects. Commonly, the $|(qc)_{\bar{3}}(\bar{q}\bar{c})_{3}\rangle$ and $|(qc)_{6}(\bar{q}\bar{c})_{\bar{6}}\rangle$ configurations are called the diquark-antidiquark configuration, which is illustrated in Fig. \ref{Jacobi}(a). The representations of Figs. \ref{Jacobi}(b) and (c) are called the meson-meson configuration, which is composed by the product of two color-singlets ($|(q\bar{c})_{1}(c\bar{q})_{1}\rangle$) or the
product of two color-octets ($|(q\bar{c})_{8}(c\bar{q})_{8}\rangle$). As for the meson-meson configuration, we are more interested in the configuration of Fig. \ref{Jacobi}(c). The color relations between the diquark-antidiquark and meson-meson configuration of Fig. \ref{Jacobi}(c) can be expressed as follows,
\begin{eqnarray}
|(q\bar{c})_{1}(c\bar{q})_{1}\rangle =&-\sqrt{\frac{1}{3}}|(qc)_{\bar{3}}(\bar{q}\bar{c})_{3}\rangle+\sqrt{\frac{2}{3}}|(qc)_{6}(\bar{q}\bar{c})_{\bar{6}}\rangle \\
|(q\bar{c})_{8}(c\bar{q})_{8}\rangle=&\sqrt{\frac{2}{3}}|(qc)_{\bar{3}}(\bar{q}\bar{c})_{3}\rangle+\sqrt{\frac{1}{3}}|(qc)_{6}(\bar{q}\bar{c})_{\bar{6}}\rangle
\end{eqnarray}
Basing on these relations, we also obtain its proportions in the meson-meson configuration, which are also shown in the last two columns in Tables \ref{qcqc2}$\sim$\ref{qcqc3}, \ref{scsc2}$\sim$\ref{scsc3} and \ref{qcsc2}$\sim$\ref{qcsc3}. For $1^{+-}$(3902) as an example, which is a good candidate for experimental structure Z$_{c}$(3900), it contains 44.6$\%$ $|(qc)_{1}(\bar{q}\bar{c})_{1}\rangle$ components and 55.4$\%$ $|(qc)_{8}(\bar{q}\bar{c})_{8}\rangle$ ones.
\section{The mass spectra of hidden-charm tetraquarks}\label{sec4}
To show the mass spectra more explicitly, we depict the results in Figs. \ref{qcqcfig}-\ref{qcscfig} together with the experimental data, where black and pink solid lines denote the masses of the ground and the first radially excited states respectively, black inverted triangles represent the experimental data, dashed lines are corresponding S-wave charm meson pair thresholds. From these figures, one can see that the mass gaps between the ground states and the first radial excitations are about 500$\sim$550 MeV. This behavior is similar with that of the mass spectrum of the $c\bar{c}$ system. Another important feature is that some predicted states are located very near with each other, $e$.$g$. the $1^{+}$(4182), $1^{+}$(4215) and $1^{+}$(4251) states of $qc\bar{s}\bar{c}$ system(see Fig. \ref{qcscfig}). If the mass splitting of these states is smaller than their decay widths, they will overlap with each other and contribute to a broad structure in the invariant spectrum. The Z$_{cs}$(4220) structure with its mass and width to be ($4216\pm24$$^{+43}_{-30}$) MeV and ($233\pm52$$^{+97}_{-73}$) MeV, may arise from these $1^{+}$ $qc\bar{s}\bar{c}$ states.

\subsection{The $qc\bar{q}\bar{c}$ ($q=u/d$) system}\label{sec4-1}

The famous $X(3872)$ state was firstly observed by Belle collaboration in 2003, its quantum number, mass and width were determined to be $J^{PC}=1^{++}$, $M=(3872\pm0.6\pm0.5$) MeV and $\Gamma<2.3$ MeV. From Fig. \ref{qcqcfig}, it is shown that the lowest energy of the $1^{++}$ $qc\bar{q}\bar{c}$ system is 3974 MeV which is much higher than the experimental data of $X(3872)$. Thus, the present work do not support $X(3872)$ to be a compact tetraquark state. The charged charmonium-like state $Z^{+}(4430)$ was also discovered by Belle Collaboration with its measured mass being ($4433\pm4\pm2$) MeV. This value is consistent well with the $1^{+-}$(4428) state which is the first radial excitation of $qc\bar{q}\bar{c}$ system(see Fig. \ref{qcqcfig}). $Z_{c}(3900)$ has been supposed to be the ground state with $J^{PC}=1^{+-}$, therein, $Z^{+}(4430)$ can be assigned as the first radially excited state of $Z_{c}$(3900). In 2008, the Belle collaboration reported two resonance-like structures $X(4050)$ and $X(4250)$ in the $\pi^{+}\chi_{c1}$ invariant mass distribution, and determined their masses $M_{X(4050)}=(4051\pm14^{+20}_{-41})$ MeV, $M_{X(4250)}$=($4248^{+44}_{-29}$$^{+180}_{-35}$) MeV, respectively. In the present model, there are two $qc\bar{q}\bar{c}$ states $0^{++}$(4085) and $0^{++}$(4255) whose energies are close to the experimental data of $X(4050)$ and $X(4250)$. However, the quantum numbers of these structures have not been determined in experiments, evenly, the existence of them still needs confirmation by other experimental collaborations.

Another charged charmonium-like structure $Z_{c}(4200)$ \cite{Belle4200} was observed by Belle collaboration with its spin-parity being suggested to be $J^{P}=1^{+}$. The measured mass and width of this state are ($4196^{+31}_{-29}$$^{+17}_{-13}$) MeV and ($370^{+70}_{-70}$$^{+70}_{-132}$) MeV, respectively. In the present work, there is a predicted $1^{+-}$ state with its energy being 4155 MeV which is about 40 MeV lower than the mass of $Z_{c}(4200)$. Considering the broad structure of $Z_{c}(4200)$, it can be tentatively assigned as a $1^{+}$ compact tetraquark state.

The $Z(3930)$ structure was observed in the $D\overline{D}$ invariant mass spectrum in the process $\gamma\gamma\rightarrow D\overline{D}$, with the mass $M=(3929\pm5\pm2)$ MeV, width $\Gamma=(29\pm10\pm2)$ MeV. Because this state is produced according to $\gamma\gamma$ fusion process, its quantum numbers should be $J^{PC}=0^{++}$ or $2^{++}$. The predicted energy of the next lowest state of $0^{++}$ $qc\bar{q}\bar{c}$ system is 3947 MeV which is very close to the measured mass of $Z(3930)$. Therefore, $Z(3930)$ can be described as a compact hidden-charm tetraquark with $J^{PC}=0^{++}$.

The $Z_{c}(4025)$ structure was first discovered in the $D^{*}\bar{D}^{*}$ mass spectrum by BESIII Collaboration. Almost at the same time they observed another charged charmonium-like state which was named as $Z_{c}(4020)$, in the $\pi^{\pm}h_{c}$ invariant mass distribution. At present, these two structures are denoted as the same state $X(4020)$ in PDG with spin-parity $J^{P}=1^{+}$. Its mass and width are $M=(4022.9\pm0.8\pm2.7$) MeV, $\Gamma=(7.9\pm2.7\pm2.6$) MeV. It is shown in Fig. \ref{qcqcfig}, the closest state of theoretical prediction in energy to $X(4020)$ is the lowest $1^{++}$ state with $M=3974$ MeV which is more than 40 MeV lower than the experimental data of $X(4020)$. Therefore, whether $X(4020)$ can be described as a hidden-charm tetraquark needs further confirmations in experiments and theories.
\subsection{The $sc\bar{s}\bar{c}$ system}\label{sec4-2}
In 2009, CDF Collaboration announced a narrow structure $Y(4140)$ near the $J/\psi \phi$ threshold. It was discovered in the exclusive $B\rightarrow KJ/\psi\phi$ decay with its mass and width measured to be $M=(4143.0\pm2.9\pm1.2)$ MeV  and $\Gamma=(11.7^{+8.3}_{-5.0}\pm3.7)$ MeV, respectively. Later in 2011, the CDF Collaboration confirmed the $Y(4140)$ by performing a further study based on the increased $B^{+}\rightarrow J/\psi\phi K^{+}$ sample. Besides, they also reported another structure, named as $Y(4274)$, in the $J/\psi\phi$ invariant mass spectrum with mass and width being
$M=(4274.4^{+8.4}_{-6.7}\pm1.9)$ MeV and $\Gamma=(32.3^{+21.9}_{-15.3}\pm0.097)$ MeV. Since these two resonances were both observed in the $J/\psi\phi$ decay mode, their C-parity and G-parity should be even. In present work, we also obtain two $sc\bar{s}\bar{c}$ states with their energies very close to the measured masses of $Y(4140)$ and $Y(4274)$ respectively(see Fig. \ref{scscfig}). They are the two ground $1^{++}$ states with their theoretical masses to be $4147$ and $4274$ MeV. Thus, the present work support assigning $Y(4140)$ and $Y(4274)$ as compact tetrquarks with quark components $sc\bar{s}\bar{c}$. Besides of $Y(4140)$ and $Y(4274)$, another new narrow structure named as $X(4350)$ \cite{Belle4350}, which was also detected in the $\phi J/\psi$ invariant mass spectrum, can also be interpreted as a $sc\bar{s}\bar{c}$ tetraquark. The measured mass of $X(4350)$ is $M=(4350.6^{+4.6}_{-5.1}\pm0.7)$ MeV. It can be seen from Fig. \ref{scscfig} this value is compatible with the theoretical mass of $0^{++}$(4335). This indicates $0^{++}$(4335) state may be a possible assignment for $X(4350)$.

Another two extra structures with higher masses were found in the $J/\psi\phi$ invariant mass spectrum, which are $X(4500)$ \cite{LHCb4500} and $X(4700)$ \cite{LHCb4700} with $J^{P}=0^{+}$. Their masses and widths are measured to be $M_{X(4500)}=$($4506\pm11$$^{+12}_{-15}$) MeV, $\Gamma_{X(4500)}=(92\pm21^{+21}_{-20})$ MeV and $M_{X(4700)}=$($4704\pm10$$^{+14}_{-24}$) MeV, $\Gamma_{X(4700)}=(120\pm31^{+42}_{-33})$ MeV. From Fig. \ref{scscfig}, it is shown that the most possible assignments for these two structures are $0^{++}$(4466) and $0^{++}$(4753) states. We can see that the theoretical predictions are not consistent well with the experimental data. Considering the decay widths and the model uncertainties, theoretical values are roughly compatible with the experimental data. We look forward to further confirmations about the natures of these two structures in experiments and theories.
\subsection{The $qc\bar{s}\bar{c}$ ($q=u/d$) system}\label{sec4-3}
Finally, we turn to the $Z_{cs}$ states which have the quark components of $qc\bar{s}\bar{c}$. In 2021, a charmonium-like state $Z_{cs}(3985)$ containing a strange quark was discovered in the $D_{s}D^{*}$ channel by BESIII. Its mass and width are determined to be $M=(3982.5^{+1.8}_{-2.6}\pm2.1)$ MeV and $\Gamma=(12.8^{+5.3}_{-4.4}\pm3.0)$ MeV. Besides, LHCb Collaboration reported two $Z_{cs}$ particles, $Z_{cs}(4000)$ and $Z_{cs}(4220)$ in the $J/\psi K^{+}$ mass distributions. The spin-parity was suggested to be $J^{P}=1^{+}$ for $Z_{cs}(4000)$, and $1^{+}$ or $1^{-}$ for $Z_{cs}(4220)$. Their measured masses and widths are
\begin{eqnarray}
\notag
&&M_{Z_{cs}(4000)}=4003\pm6^{+4}_{-14}   \quad  \mathrm{MeV},  \\
\notag
&&\Gamma_{Z_{cs}(4000)}=131\pm15\pm26   \quad   \mathrm{MeV},  \\
\notag
&&M_{Z_{cs}(4220)}=4216\pm24^{+43}_{-30} \quad  \mathrm{MeV},     \\
\notag
&&\Gamma_{Z_{cs}(4220)}=233\pm52^{+97}_{-73} \quad  \mathrm{MeV}
\end{eqnarray}
It can be seen from Fig. \ref{qcscfig} that the broad structure $Z_{cs}(4220)$ fall in the energy range of 4182$\sim$4251 MeV with $J^{P}=1^{+}$. Therefore, we speculate that a compact tetraquark state with $J^{P}=1^{+}$ may take considerable probability for $Z_{cs}(4220)$. In addition, the lowest energy of $J^{P}=1^{+}$ state is 4034 MeV which is roughly compatible with the mass of $Z_{cs}(4000)$. Basing on this results, $Z_{cs}(4000)$ structure may be also a compact tetraquark and be a partner of $Z_{cs}(4220)$. As for the $Z_{cs}(3985)$ structure, it is shown by the mass spectrum in Fig. \ref{qcscfig} that no matching state consist for it. Thus, $Z_{cs}(3985)$ is disfavored to be a compact tetraquark state at present.

\begin{large}
\section{Conclusions}\label{sec5}
\end{large}

In the present work, we have systematically studied the mass spectra, the r.m.s. radii and the radial density distributions of the ground states and the first radially excited states of $qc\overline{q}\bar{c}$ ($q$=$u/d$ or $s$ quark) systems. The calculation is carried out in the frame work of relativized quark model, where the Coulomb term, confining potential, tensor potential and contact interaction are all considered. In the first stage, the masses, r.m.s. radii, and radial density distributions of different color configurations are calculated. Then, we obtain the mass spectra and r.m.s. radii of the physical states by considering the mixing effect.

According to the results, we find some interesting phenomenons. For example, $|(qc)_{6}(\bar{q}\bar{c})_{\bar{6}}\rangle$ configuration commonly has higher energy than that of $|(qc)_{\bar{3}}(\bar{q}\bar{c})_{3}\rangle$. In addition, the r.m.s. radii $\sqrt{\langle r_{12/34}^{2}\rangle}$ of $|(qc)_{\bar{3}}(\bar{q}\bar{c})_{3}\rangle$ configuration are smaller than those of the $|(qc)_{6}(\bar{q}\bar{c})_{\bar{6}}\rangle$, while the situation is exactly opposite to $\sqrt{\langle r^{2}\rangle}$. Finally, it is shown that the r.m.s. radii of hidden-charm tetraquarks are smaller than 1 fm, which implies that two charmed and two light quarks have large possibility constituting compact tetraquark states.

The numerical results show that predicted energies of several compact tetraquark states are very close to the masses of experimentally observed states. Based on these results, some potential candidates for hidden-charm tetraquark states are suggested. If assigning $Z_{c}(3900)$ as the ground state of $qc\bar{q}\bar{c}$($q=u/d$) system with $J^{PC}=1^{+-}$, theoretical predictions support identifying $Z(4430)$ to be the first radially excited states of $Z_{c}(3900)$. The broad structure $Z_{c}(4200)$ can also be tentatively interpreted as a partner of $Z_{c}(3900)$ with $J^{PC}=1^{+-}$. Besides, the possible assignments for $X(3930)$, $X(4050)$ and $X(4250)$ structures are low-lying $0^{++}$ states with $qc\bar{q}\bar{c}$ ($q=u/d$) contents. As for the $sc\bar{s}\bar{c}$ system, theoretical predictions indicate that $X(4140)$ and $X(4274)$ structures have much possibilities to be this type of tetraquarks with $J^{PC}=1^{++}$. In addition, $X(4350)$ structure can also be described as a $sc\bar{s}\bar{c}$ tetraquark with $J^{PC}=0^{++}$. With regard to $qc\bar{s}\bar{c}$ ($q=u/d$) system, we find two potential candidates for this type of tetraquarks, they are the $Z_{cs}(4000)$ and $Z_{cs}(4220)$ structures which can be described as the $J^{P}=1^{+}$ states. It is noted that these above assignments and suggestions are proposed only according to their mass spectra, which needs further confirmations. The final conclusions about the nature of these exotic states should be determined by the mass spectra together with their decay properties and production processes.

\section*{Acknowledgements}

This project is supported by National Natural Scieqce Foundation, Grant Number 12175068 and Natural Science Foundation of HeBei Province, Grant Number A2024502002.

\begin{widetext}
	\begin{large}
		\textbf{Appendix: The mass matrix and the results of diagonalizing the mass matrix.}
	\end{large}
\begin{table*}[htbp]
\begin{ruledtabular}\caption{The mass matrix (MeV), the eigenvalue (MeV) and the eigenvector for the ground state of $qc\bar{q}\bar{c}$($q=u/d$) system by diagonalizing the mass matrix.}
\label{qcqcA1}
\begin{tabular}{c| c| c |c c   }
\multirow{2}{*}{$J^{PC}$ }&  \multirow{2}{*}{Configuration } & \multicolumn{3}{c}{Configuration mixing(I)}  \\ \cline{3-5}
\multirow{2}{*}{ }        &    \multirow{2}{*}{ }   & $ H$     & Eigenvalue &   Eigenvector \\ \hline
\multirow{4}{*}{$0^{++}$}& $|(qc)_{1}^{\bar{3}}(\bar{q}\bar{c})_{1}^{3}\rangle_{0}$ & \multirow{4}{*}{$\begin{pmatrix} 4038 & -21 & -62& -141\\ -21 & 3970 &-163 & 0  \\ -62 & -163& 3834&-65\\ -141 & 0 &-65 & 4102\\\end{pmatrix}$} & 4217 & (-0.605, 0.104, -0.079, 0.785)   \\
\multirow{4}{*}{ }&           $|(qc)_{0}^{\bar{3}}(\bar{q}\bar{c})_{0}^{3}\rangle_{0}$ & \multirow{4}{*}{}      & 4085 & (0.272, 0.768, -0.578, 0.0504)  \\
\multirow{4}{*}{ }&           $|(qc)_{1}^{6}(\bar{q}\bar{c})_{1}^{\bar{6}}\rangle_{0}$ & \multirow{4}{*}{}      & 3947 & (0.700, -0.395, -0.145, 0.577)  \\
\multirow{4}{*}{ }&           $|(qc)_{0}^{6}(\bar{q}\bar{c})_{0}^{\bar{6}}\rangle_{0}$ & \multirow{4}{*}{}      & 3695 & (-0.264, -0.494, -0.799, -0.219)  \\  \hline
\multirow{2}{*}{$1^{++}$}&  $\frac{1}{\sqrt{2}}$[$|(qc)_{1}^{\bar{3}}(\bar{q}\bar{c})_{0}^{3}\rangle_{1}$+$|(qc)_{0}^{\bar{3}}(\bar{q}\bar{c})_{1}^{3}\rangle_{1}$] & \multirow{2}{*}{$\begin{pmatrix} 4039 & 82 \\ 82 & 4077 \end{pmatrix}$} & 4142 &   (0.622,0.783)   \\
\multirow{2}{*}{ }&  $\frac{1}{\sqrt{2}}$[$|(qc)_{1}^{6}(\bar{q}\bar{c})_{0}^{\bar{6}}\rangle_{1}$+$|(qc)_{0}^{6}(\bar{q}\bar{c})_{1}^{\bar{6}}\rangle_{1}$] &  & 3974 &  (-0.783,0.622)   \\  \hline
\multirow{4}{*}{$1^{+-}$}& $\frac{1}{\sqrt{2}}$[$|(qc)_{1}^{\bar{3}}(\bar{q}\bar{c})_{0}^{3}\rangle_{1}$-$|(qc)_{0}^{\bar{3}}(\bar{q}\bar{c})_{1}^{3}\rangle_{1}$] & \multirow{4}{*}{$\begin{pmatrix} 4039 & -82 &30 & 61\\ -82 & 4077 &58 & 54\\ 30& 58 &4076 & -27\\  61&  54& -27& 4066\\\end{pmatrix}$} & 4155 & (0.427, -0.805, -0.406, -0.073)   \\
\multirow{4}{*}{ }&           $\frac{1}{\sqrt{2}}$[$|(qc)_{1}^{6}(\bar{q}\bar{c})_{0}^{\bar{6}}\rangle_{1}$-$|(qc)_{0}^{6}(\bar{q}\bar{c})_{1}^{\bar{6}}\rangle_{1}$] & \multirow{4}{*}{}      & 4113 & (0.532, 0.164, 0.086, 0.826)  \\
\multirow{4}{*}{ }&           $|(qc)_{1}^{\bar{3}}(\bar{q}\bar{c})_{1}^{3}\rangle_{1}$ & \multirow{4}{*}{}      & 4089 & (0.410, -0.176, 0.838, -0.315)  \\
\multirow{4}{*}{ }&           $|(qc)_{1}^{6}(\bar{q}\bar{c})_{1}^{\bar{6}}\rangle_{1}$ & \multirow{4}{*}{}      & 3902 & (0.606, 0.542, -0.356, -0.461)  \\  \hline
\multirow{2}{*}{$2^{++}$}&  $|(qc)_{1}^{\bar{3}}(\bar{q}\bar{c})_{1}^{3}\rangle_{2}$ & \multirow{2}{*}{$\begin{pmatrix} 4142 & 25 \\ 25 & 4147 \end{pmatrix}$} & 4170 &   (0.671, 0.741)   \\
\multirow{2}{*}{ }&  $|(qc)_{1}^{6}(\bar{q}\bar{c})_{1}^{\bar{6}}\rangle_{2}$ &  & 4119 &  (-0.741, 0.671)   \\
\end{tabular}
\end{ruledtabular}
\end{table*}
\begin{table*}[htbp]
\begin{ruledtabular}\caption{Same as in TABLE XI for the firs radially excited states of $qc\bar{q}\bar{c}$($q=u/d$) system.}
\label{qcqcA2}
\begin{tabular}{c| c| c |c c   }
\multirow{2}{*}{$J^{PC}$ }&  \multirow{2}{*}{Configuration } & \multicolumn{3}{c}{Configuration mixing(II)}  \\ \cline{3-5}
\multirow{2}{*}{ }        &    \multirow{2}{*}{ }   & $ H$     & Eigenvalue &   Eigenvector \\ \hline
\multirow{4}{*}{$0^{++}$}& $|(qc)_{1}^{\bar{3}}(\bar{q}\bar{c})_{1}^{3}\rangle_{0}$ & \multirow{4}{*}{$\begin{pmatrix} 4559 & -15 &-58 & -147\\ -15 & 4482 & -172& 0 \\ -58 & -172& 4425&-51\\-147  & 0 & -51& 4635\\\end{pmatrix}$} & 4750 & (-0.601, 0.068, -0.0533, 0.794)   \\
\multirow{4}{*}{ }&           $|(qc)_{0}^{\bar{3}}(\bar{q}\bar{c})_{0}^{3}\rangle_{0}$ & \multirow{4}{*}{}      & 4636 & (0.228, 0.713, -0.659, 0.067)  \\
\multirow{4}{*}{ }&           $|(qc)_{1}^{6}(\bar{q}\bar{c})_{1}^{\bar{6}}\rangle_{0}$ & \multirow{4}{*}{}      & 4460 & (-0.717, 0.388, 0.113, -0.568)  \\
\multirow{4}{*}{ }&           $|(qc)_{0}^{6}(\bar{q}\bar{c})_{0}^{\bar{6}}\rangle_{0}$ & \multirow{4}{*}{}      & 4255 & (-0.269, -0.580, -0.741, -0.204)  \\  \hline
\multirow{2}{*}{$1^{++}$}&  $\frac{1}{\sqrt{2}}$[$|(qc)_{1}^{\bar{3}}(\bar{q}\bar{c})_{0}^{3}\rangle_{1}$+$|(qc)_{0}^{\bar{3}}(\bar{q}\bar{c})_{1}^{3}\rangle_{1}$] & \multirow{2}{*}{$\begin{pmatrix} 4546 &  90\\  90& 4615 \end{pmatrix}$} & 4677 &   (0.567, 0.824)   \\
\multirow{2}{*}{ }&  $\frac{1}{\sqrt{2}}$[$|(qc)_{1}^{6}(\bar{q}\bar{c})_{0}^{\bar{6}}\rangle_{1}$+$|(qc)_{0}^{6}(\bar{q}\bar{c})_{1}^{\bar{6}}\rangle_{1}$] &  & 4484 &  (-0.824, 0.567)   \\  \hline
\multirow{4}{*}{$1^{+-}$}& $\frac{1}{\sqrt{2}}$[$|(qc)_{1}^{\bar{3}}(\bar{q}\bar{c})_{0}^{3}\rangle_{1}$-$|(qc)_{0}^{\bar{3}}(\bar{q}\bar{c})_{1}^{3}\rangle_{1}$] & \multirow{4}{*}{$\begin{pmatrix} 4546 & -90 & 21& 64\\ -90 & 4615 &59 & 35\\ 21& 59 &4579 & -28\\  64& 35 & -28& 4606\\\end{pmatrix}$} & 4692 & (0.500, -0.773, -0.352, 0.171)   \\
\multirow{4}{*}{ }&           $\frac{1}{\sqrt{2}}$[$|(qc)_{1}^{6}(\bar{q}\bar{c})_{0}^{\bar{6}}\rangle_{1}$-$|(qc)_{0}^{6}(\bar{q}\bar{c})_{1}^{\bar{6}}\rangle_{1}$] & \multirow{4}{*}{}      & 4639 & (0.274, 0.360, 0.034, 0.891)  \\
\multirow{4}{*}{ }&           $|(qc)_{1}^{\bar{3}}(\bar{q}\bar{c})_{1}^{3}\rangle_{1}$ & \multirow{4}{*}{}      & 4587 & (0.478, -0.113, 0.861, -0.134)  \\
\multirow{4}{*}{ }&           $|(qc)_{1}^{6}(\bar{q}\bar{c})_{1}^{\bar{6}}\rangle_{1}$ & \multirow{4}{*}{}      & 4428 & (0.669, 0.511, -0.366, -0.398)  \\  \hline
\multirow{2}{*}{$2^{++}$}&  $|(qc)_{1}^{\bar{3}}(\bar{q}\bar{c})_{1}^{3}\rangle_{2}$ & \multirow{2}{*}{$\begin{pmatrix} 4633 & 27 \\ 27 & 4670 \end{pmatrix}$} & 4684 &   (0.466, 0.885)   \\
\multirow{2}{*}{ }&  $|(qc)_{1}^{6}(\bar{q}\bar{c})_{1}^{\bar{6}}\rangle_{2}$ &  & 4619 &  (-0.885, 0.466)   \\
\end{tabular}
\end{ruledtabular}
\end{table*}
\begin{table*}[htbp]
\begin{ruledtabular}\caption{Same as in TABLE XI for the ground states of $sc\bar{s}\bar{c}$ system.}
\label{scscA1}
\begin{tabular}{c| c| c |c c   }
\multirow{2}{*}{$J^{PC}$ }&  \multirow{2}{*}{Configuration } & \multicolumn{3}{c}{Configuration mixing(I)}  \\ \cline{3-5}
\multirow{2}{*}{ }        &    \multirow{2}{*}{ }   & $ H$     & Eigenvalue &   Eigenvector \\ \hline
\multirow{4}{*}{$0^{++}$}& $|(sc)_{1}^{\bar{3}}(\bar{sc})_{1}^{3}\rangle_{0}$ & \multirow{4}{*}{$\begin{pmatrix} 4210 & -9 & -28& -110\\ -9 & 4144 &-127 & 0  \\ -28 & -127& 4031&-30\\ -110 & 0 &-30 & 4237\\\end{pmatrix}$} & 4335 & (-0.657, 0.055, -0.037, 0.751)   \\
\multirow{4}{*}{ }&           $|(sc)_{0}^{\bar{3}}(\bar{sc})_{0}^{3}\rangle_{0}$ & \multirow{4}{*}{}      & 4229 & (0.155, 0.816, -0.554, 0.048)  \\
\multirow{4}{*}{ }&           $|(sc)_{1}^{6}(\bar{sc})_{1}^{\bar{6}}\rangle_{0}$ & \multirow{4}{*}{}      & 4119 & (0.720, -0.242, -0.099, 0.643)  \\
\multirow{4}{*}{ }&           $|(sc)_{0}^{6}(\bar{sc})_{0}^{\bar{6}}\rangle_{0}$ & \multirow{4}{*}{}      & 3940 & (-0.161, -0.521, -0.826, -0.143)  \\  \hline
\multirow{2}{*}{$1^{++}$}&  $\frac{1}{\sqrt{2}}$[$|(sc)_{1}^{\bar{3}}(\bar{sc})_{0}^{3}\rangle_{1}$+$|(sc)_{0}^{\bar{3}}(\bar{sc})_{1}^{3}\rangle_{1}$] & \multirow{2}{*}{$\begin{pmatrix} 4204 & 63 \\ 63 & 4217 \end{pmatrix}$} & 4274 &   (0.670, 0.743)   \\
\multirow{2}{*}{ }&  $\frac{1}{\sqrt{2}}$[$|(sc)_{1}^{6}(\bar{sc})_{0}^{\bar{6}}\rangle_{1}$+$|(sc)_{0}^{6}(\bar{sc})_{1}^{\bar{6}}\rangle_{1}$] &  & 4147 &   (-0.743, 0.670)  \\  \hline
\multirow{4}{*}{$1^{+-}$}& $\frac{1}{\sqrt{2}}$[$|(sc)_{1}^{\bar{3}}(\bar{sc})_{0}^{3}\rangle_{1}$-$|(sc)_{0}^{\bar{3}}(\bar{sc})_{1}^{3}\rangle_{1}$] & \multirow{4}{*}{$\begin{pmatrix} 4204 & -63 &15 & 25\\ -63 & 4217 &26 & 21\\ 15& 26 &4236 & -12\\  25&  21& -12& 4208\\\end{pmatrix}$} & 4276 & (0.608, -0.747, -0.266, 0.040)   \\
\multirow{4}{*}{ }&           $\frac{1}{\sqrt{2}}$[$|(sc)_{1}^{6}(\bar{sc})_{0}^{\bar{6}}\rangle_{1}$-$|(sc)_{0}^{6}(\bar{sc})_{1}^{\bar{6}}\rangle_{1}$] & \multirow{4}{*}{}      & 4242 & (0.370, -0.033, 0.925, -0.075)  \\
\multirow{4}{*}{ }&           $|(sc)_{1}^{\bar{3}}(\bar{sc})_{1}^{3}\rangle_{1}$ & \multirow{4}{*}{}      & 4222 & (0.273, 0.280, -0.025, 0.920)  \\
\multirow{4}{*}{ }&           $|(sc)_{1}^{6}(\bar{sc})_{1}^{\bar{6}}\rangle_{1}$ & \multirow{4}{*}{}      & 4124 & (0.647, 0.602, -0.268, -0.383)  \\  \hline
\multirow{2}{*}{$2^{++}$}&  $|(sc)_{1}^{\bar{3}}(\bar{sc})_{1}^{3}\rangle_{2}$ & \multirow{2}{*}{$\begin{pmatrix} 4290 & 11 \\ 11 & 4270 \end{pmatrix}$} & 4295 &   (-0.915, -0.405)   \\
\multirow{2}{*}{ }&  $|(sc)_{1}^{6}(\bar{sc})_{1}^{\bar{6}}\rangle_{2}$ &  & 4265 &  (0.405, -0.915)  \\
\end{tabular}
\end{ruledtabular}
\end{table*}
\begin{table*}[htbp]
\begin{ruledtabular}\caption{Same as in TABLE XI for the firs radially excited states of $sc\bar{s}\bar{c}$ system.}
\label{scscA2}
\begin{tabular}{c| c| c |c c   }
\multirow{2}{*}{$J^{PC}$ }&  \multirow{2}{*}{Configuration } & \multicolumn{3}{c}{Configuration mixing(II)}  \\ \cline{3-5}
\multirow{2}{*}{ }        &    \multirow{2}{*}{ }   & $ H$     & Eigenvalue &   Eigenvector \\ \hline
\multirow{4}{*}{$0^{++}$}& $|(sc)_{1}^{\bar{3}}(\bar{sc})_{1}^{3}\rangle_{0}$ & \multirow{4}{*}{$\begin{pmatrix} 4719 & -7 & -27& -116\\ -7 & 4647 &-134 & 0  \\ -27 & -134& 4576&-26\\ -116 & 0 &-26 & 4748\\\end{pmatrix}$} & 4851 & (-0.659, 0.039, -0.026, 0.751)   \\
\multirow{4}{*}{ }&           $|(sc)_{0}^{\bar{3}}(\bar{sc})_{0}^{3}\rangle_{0}$ & \multirow{4}{*}{}      & 4753 & (0.136, 0.774, -0.616, 0.057)  \\
\multirow{4}{*}{ }&           $|(sc)_{1}^{6}(\bar{sc})_{1}^{\bar{6}}\rangle_{0}$ & \multirow{4}{*}{}      & 4621 & (-0.722, 0.242, 0.084, -0.643)  \\
\multirow{4}{*}{ }&           $|(sc)_{0}^{6}(\bar{sc})_{0}^{\bar{6}}\rangle_{0}$ & \multirow{4}{*}{}      & 4466 & (-0.163, -0.585, -0.783, -0.139)  \\  \hline
\multirow{2}{*}{$1^{++}$}&  $\frac{1}{\sqrt{2}}$[$|(sc)_{1}^{\bar{3}}(\bar{sc})_{0}^{3}\rangle_{1}$+$|(sc)_{0}^{\bar{3}}(\bar{sc})_{1}^{3}\rangle_{1}$] & \multirow{2}{*}{$\begin{pmatrix} 4702 & 71 \\ 71 & 4731 \end{pmatrix}$} & 4784 &   (0.653,0.757)   \\
\multirow{2}{*}{ }&  $\frac{1}{\sqrt{2}}$[$|(sc)_{1}^{6}(\bar{sc})_{0}^{\bar{6}}\rangle_{1}$+$|(sc)_{0}^{6}(\bar{sc})_{1}^{\bar{6}}\rangle_{1}$] &  & 4641 &  (-0.757,0.653)   \\  \hline
\multirow{4}{*}{$1^{+-}$}& $\frac{1}{\sqrt{2}}$[$|(sc)_{1}^{\bar{3}}(\bar{sc})_{0}^{3}\rangle_{1}$-$|(sc)_{0}^{\bar{3}}(\bar{sc})_{1}^{3}\rangle_{1}$] & \multirow{4}{*}{$\begin{pmatrix} 4702 & -71 &10 & 27\\ -71 & 4731 &45 & 16\\ 10& 45 &4731 & -27\\  27&  16& -27& 4723\\\end{pmatrix}$} & 4804 & (0.505, -0.723, -0.439, 0.172)   \\
\multirow{4}{*}{ }&           $\frac{1}{\sqrt{2}}$[$|(sc)_{1}^{6}(\bar{sc})_{0}^{\bar{6}}\rangle_{1}$-$|(sc)_{0}^{6}(\bar{sc})_{1}^{\bar{6}}\rangle_{1}$] & \multirow{4}{*}{}      & 4743 & (-0.256, 0.335, -0.569, 0.706)  \\
\multirow{4}{*}{ }&           $|(sc)_{1}^{\bar{3}}(\bar{sc})_{1}^{3}\rangle_{1}$ & \multirow{4}{*}{}      & 4724 & (0.524, 0.148, 0.591, 0.596)  \\
\multirow{4}{*}{ }&           $|(sc)_{1}^{6}(\bar{sc})_{1}^{\bar{6}}\rangle_{1}$ & \multirow{4}{*}{}      & 4616 & (0.637, 0.586, -0.366, -0.342)  \\  \hline
\multirow{2}{*}{$2^{++}$}&  $|(sc)_{1}^{\bar{3}}(\bar{sc})_{1}^{3}\rangle_{2}$ & \multirow{2}{*}{$\begin{pmatrix} 4775 & 12 \\ 12 & 4776 \end{pmatrix}$} & 4788 &   (0.692, 0.722)   \\
\multirow{2}{*}{ }&  $|(sc)_{1}^{6}(\bar{sc})_{1}^{\bar{6}}\rangle_{2}$ &  & 4763 &  ( -0.722,0.692)   \\
\end{tabular}
\end{ruledtabular}
\end{table*}
\begin{table*}[htbp]
\begin{ruledtabular}\caption{Same as in TABLE XI for the ground states of $qc\bar{s}\bar{c}$($q=u/d$) system.}
\label{qcscA1}
\begin{tabular}{c| c| c |c c   }
\multirow{2}{*}{$J^{PC}$ }&  \multirow{2}{*}{Configuration } & \multicolumn{3}{c}{Configuration mixing(I)}  \\ \cline{3-5}
\multirow{2}{*}{ }        &    \multirow{2}{*}{ }   & $ H$     & Eigenvalue &   Eigenvector \\ \hline
\multirow{4}{*}{$0^{+}$}& $|(qc)_{1}^{\bar{3}}(\bar{sc})_{1}^{3}\rangle_{0}$ & \multirow{4}{*}{$\begin{pmatrix} 4129 & -12 & -38& -120\\ -12 & 4059 &-138 & 0  \\ -38 & -138& 3946&-41\\ -120 & 0 &-41 & 4174\\\end{pmatrix}$} & 4275 & (-0.630, 0.069, -0.052, 0.773)   \\
\multirow{4}{*}{ }&           $|(qc)_{0}^{\bar{3}}(\bar{sc})_{0}^{3}\rangle_{0}$ & \multirow{4}{*}{}      & 4155 & (0.204, 0.793, -0.571, 0.057)  \\
\multirow{4}{*}{ }&           $|(qc)_{1}^{6}(\bar{sc})_{1}^{\bar{6}}\rangle_{0}$ & \multirow{4}{*}{}      & 4039 & (0.723, -0.307, -0.108, 0.609)  \\
\multirow{4}{*}{ }&           $|(qc)_{0}^{6}(\bar{sc})_{0}^{\bar{6}}\rangle_{0}$ & \multirow{4}{*}{}      & 3840 & (-0.199, -0.521, -0.812, -0.171)  \\  \hline
\multirow{6}{*}{$1^{+}$}&  $|(qc)_{1}^{\bar{3}}(\bar{sc})_{0}^{3}\rangle_{1}$ & \multirow{6}{*}{$\begin{pmatrix} 4130 & 7 & 0& 70&-5 &27\\ 7 & 4118 &70 & 0& -39& 58\\ 0 & 70 &4149 &21 &28 &21\\ 70 & 0 & 21& 4156& -24&-25\\ -5 & -39 &28& -24& 4160&-22\\ 27 & 58 & 21&-25 &-22 &4142\end{pmatrix}$} & 4251 &   (0.358, 0.534, 0.440, 0.320, -0.305, 0.445)   \\
\multirow{6}{*}{ }&  $|(qc)_{0}^{\bar{3}}(\bar{sc})_{1}^{3}\rangle_{1}$ &  & 4215 &   (0.492, -0.308, -0.270, 0.679, -0.126, -0.335)  \\
\multirow{6}{*}{}& $|(qc)_{1}^{6}(\bar{sc})_{0}^{\bar{6}}\rangle_{1}$ &  & 4182 &  (0.083, -0.025, 0.573, 0.191, 0.762, -0.218)  \\
\multirow{6}{*}{ }&           $|(qc)_{0}^{6}(\bar{sc})_{1}^{\bar{6}}\rangle_{1}$ & \multirow{4}{*}{}      & 4131 & (0.536, -0.182, -0.337, -0.242, 0.415, 0.579)  \\
\multirow{6}{*}{ }&           $|(qc)_{1}^{\bar{3}}(\bar{sc})_{1}^{3}\rangle_{1}$ & \multirow{4}{*}{}      & 4051 & (0.559, 0.300, 0.048, -0.537, -0.107, -0.543)  \\
\multirow{6}{*}{ }&           $|(qc)_{1}^{6}(\bar{sc})_{1}^{\bar{6}}\rangle_{1}$ & \multirow{4}{*}{}      & 4034 & (0.152, -0.705, 0.538, -0.228, -0.356, 0.111)  \\  \hline
\multirow{2}{*}{$2^{+}$}&  $|(qc)_{1}^{\bar{3}}(\bar{sc})_{1}^{3}\rangle_{2}$ & \multirow{2}{*}{$\begin{pmatrix} 4217 & 15 \\ 15 & 4210 \end{pmatrix}$} & 4299 &  (-0.783, -0.622)   \\
\multirow{2}{*}{ }&  $|(qc)_{1}^{6}(\bar{sc})_{1}^{\bar{6}}\rangle_{2}$ &  & 4198 &  ( 0.622,-0.783)   \\
\end{tabular}
\end{ruledtabular}
\end{table*}
\begin{table*}[htbp]
\begin{ruledtabular}\caption{Same as in TABLE XI for the firs radially excited states of $qc\bar{s}\bar{c}$($q=u/d$) system.}
\label{qcscA2}
\begin{tabular}{c| c| c |c c   }
\multirow{2}{*}{$J^{PC}$ }&  \multirow{2}{*}{Configuration } & \multicolumn{3}{c}{Configuration mixing(II)}  \\ \cline{3-5}
\multirow{2}{*}{ }        &    \multirow{2}{*}{ }   & $ H$     & Eigenvalue &   Eigenvector \\ \hline
\multirow{4}{*}{$0^{+}$}& $|(qc)_{1}^{\bar{3}}(\bar{sc})_{1}^{3}\rangle_{0}$ & \multirow{4}{*}{$\begin{pmatrix} 4643 & -9 & -36& -126\\ -9 & 4567 &-146 & 0  \\ -36 & -146& 4512&-33\\ -126 & 0 &-33 & 4697\\\end{pmatrix}$} & 4799 & (-0.623, 0.046, -0.035, 0.780)   \\
\multirow{4}{*}{ }&           $|(qc)_{0}^{\bar{3}}(\bar{sc})_{0}^{3}\rangle_{0}$ & \multirow{4}{*}{}      & 4692 & (0.172, 0.741, -0.646, 0.065)  \\
\multirow{4}{*}{ }&           $|(qc)_{1}^{6}(\bar{sc})_{1}^{\bar{6}}\rangle_{0}$ & \multirow{4}{*}{}      & 4548 & (-0.736, 0.297, 0.085, -0.602)  \\
\multirow{4}{*}{ }&           $|(qc)_{0}^{6}(\bar{sc})_{0}^{\bar{6}}\rangle_{0}$ & \multirow{4}{*}{}      & 4380 & (-0.200, -0.601, -0.758, -0.158)  \\  \hline
\multirow{6}{*}{$1^{+}$}&  $|(qc)_{1}^{\bar{3}}(\bar{sc})_{0}^{3}\rangle_{1}$ & \multirow{6}{*}{$\begin{pmatrix} 4631 & 5 &0 & 77&56 &28\\ 5 & 4620 &77 &0 &-79 & 64\\ 0 & 77 &4676 & 18& 28&14\\ 77 & 0 & 18& 4681&-25 &-17\\ 56 & -79 &28&-25 & 4664&-40\\  28&  64& 14&-17 &-40 &4670\end{pmatrix}$} & 4788 &  (-0.028, 0.595, 0.363, 0.074, -0.492, 0.516)   \\
\multirow{6}{*}{ }&  $|(qc)_{0}^{\bar{3}}(\bar{sc})_{1}^{3}\rangle_{1}$ &  & 4746 &  (-0.595, -0.017, -0.308, -0.680, -0.296, 0.018)   \\
\multirow{6}{*}{}& $|(qc)_{1}^{6}(\bar{sc})_{0}^{\bar{6}}\rangle_{1}$ &  & 4698 & (0.084, -0.102, -0.676, 0.475, -0.548, 0.007)   \\
\multirow{6}{*}{ }&          $|(qc)_{0}^{6}(\bar{sc})_{1}^{\bar{6}}\rangle_{1}$ & \multirow{4}{*}{}      & 4661 & (0.415, -0.094, -0.372, -0.300, 0.290, 0.712)  \\
\multirow{6}{*}{ }&           $|(qc)_{1}^{\bar{3}}(\bar{sc})_{1}^{3}\rangle_{1}$ & \multirow{4}{*}{}      & 4551 & (0.538, 0.548, -0.214, -0.366, -0.079, -0.474)  \\
\multirow{6}{*}{ }&           $|(qc)_{1}^{6}(\bar{sc})_{1}^{\bar{6}}\rangle_{1}$ & \multirow{4}{*}{}      & 4498 & (-0.421, 0.571, -0.363, 0.289, 0.529, 0.037)  \\  \hline
\multirow{2}{*}{$2^{+}$}&  $|(qc)_{1}^{\bar{3}}(\bar{sc})_{1}^{3}\rangle_{2}$ & \multirow{2}{*}{$\begin{pmatrix} 4705 & 17 \\ 17 & 4726 \end{pmatrix}$} & 4735 &   (0.487, 0.873)   \\
\multirow{2}{*}{ }&  $|(qc)_{1}^{6}(\bar{sc})_{1}^{\bar{6}}\rangle_{2}$ &  & 4696 &  ( -0.873,0.487)   \\
\end{tabular}
\end{ruledtabular}
\end{table*}
\end{widetext}

\begin{thebibliography}{99}
\bibitem{qmodel2}
G. Zweig,
``Observation of a narrow charmonium-like state in exclusive $B^\pm \to K^\pm \pi^+ \pi^- J/\psi$ decays,''
An SU(3) model for strong interaction symmetry and its breaking. Version 1
\end{thebibliography}
\end{document}